\newif\iftr
\newif\ifconf
\newif\ifnonbd

\trtrue
\conffalse
\nonbdtrue

\newif\ifMORE
\MOREfalse

\ifconf
\documentclass[sigconf,review]{acmart}
\renewcommand\footnotetextcopyrightpermission[1]{} 
\else
\documentclass[sigconf,nonacm]{acmart}
\fi

\setcopyright{acmlicensed}
\copyrightyear{2026}
\acmYear{2026}
\acmDOI{XXXXXXX.XXXXXXX}
\acmConference[KDD '26]{32nd ACM SIGKDD Conference on Knowledge Discovery and Data Mining}{August 09--13,
  2026}{Jeju, Korea}
\acmISBN{978-1-4503-XXXX-X/2018/06}

\newcommand{\schemename}[0]{graph--1}
\newcommand{\schemenameS}[0]{graph--1\ }

\newcommand{\schemenameA}[0]{g1}
\newcommand{\schemenameAS}[0]{g1\ }

\renewcommand{\schemename}[0]{GraphSeek}
\renewcommand{\schemenameS}[0]{GraphSeek\ }

\renewcommand{\schemenameA}[0]{GraphSeek}
\renewcommand{\schemenameAS}[0]{GraphSeek\ }

\newcommand{\mytitle}{\schemenameA: Next-Generation Graph Analytics with LLMs}

\newif\ifcnf   
\newif\ifsq     

\newif\ifsqCAP
\newif\ifsqVS
\newif\ifsqEN
\newif\ifsqTIT

\newcommand{\ignore}[1]{}

\sqtrue
\sqCAPtrue
\sqENtrue
\sqVStrue
\sqTITtrue

\usepackage{etex}
\usepackage{epstopdf}
\usepackage{placeins}

\usepackage{graphicx}
\usepackage{float}
\usepackage{multirow}
\usepackage{rotating}
\usepackage{makecell}
\usepackage{tabulary}
\usepackage{parcolumns}
\usepackage{tikz}
\usetikzlibrary{tikzmark}

\usepackage{xpatch}
\expandafter\xpatchcmd
\csname pgfk@/tikz/every picture/.@cmd\endcsname
{\thepage}{\arabic{page}}{}{}

\tikzstyle{comment} = [draw, fill=blue!70, text=white, text width=3cm, minimum height=1cm, rounded corners, align=left, font=\scriptsize]
\tikzstyle{background_alg} = [draw, fill=blue!20, opacity=0.4, inner sep=4pt, rounded corners=2pt]

\usetikzlibrary{shapes}
\usetikzlibrary{plotmarks}
\usetikzlibrary{calc, fit}

\usepackage{enumitem}

\usepackage{amsthm}
\usepackage{amsmath,amsfonts}
\usepackage{mathtools,mathrsfs}

\usepackage{soul}
\usepackage{fontawesome}
\usepackage{pifont}
\usepackage{textcomp}
\usepackage{booktabs}
\usepackage{url}
\usepackage{pbox}
\usepackage[10pt]{moresize}

\ifsqCAP
\usepackage[font={normalfont, small}]{caption}
\usepackage[font={normalfont, small}]{subcaption}
\else
\fi

\newcommand{\vspaceSQ}[1]{\ifsqVS\vspace{#1}\fi}
\newcommand{\enlargeSQ}[1]{\ifsqEN\enlargethispage{\baselineskip}\fi}

\ifsqTIT
\ifconf
\usepackage[compact]{titlesec}
\titlespacing*{\section}{0pt}{6pt}{3pt}
\titlespacing*{\subsection}{0pt}{4pt}{2pt}
\titlespacing*{\subsubsection}{0pt}{2pt}{3pt}
\fi
\fi

\usepackage{xcolor}
\definecolor{darkgrey}{RGB}{70,70,70}
\definecolor{lightgrey}{RGB}{200,200,200}
\definecolor{lyellow}{RGB}{255,255,100}
\definecolor{llyellow}{RGB}{250,250,180}
\definecolor{lgreen}{RGB}{144,238,144}
\definecolor{raphael_comments}{RGB}{13, 145, 24}

\usepackage[customcolors]{hf-tikz}
\hfsetbordercolor{white}
\hfsetfillcolor{vlgray}

\definecolor{vlgray}{rgb}{0.77 0.77 0.77}
\definecolor{ablack}{rgb}{0.2 0.2 0.2}
\definecolor{vllgray}{rgb}{0.9 0.9 0.9}
\definecolor{bblue}{rgb}{0.7 0.7 0.99}

\usepackage{colortbl}

\usepackage{listings}

\ifsq
\else
\fi

\newcommand{\maciej}[1]{\textcolor{blue}{[Maciej: #1]}}

\newcommand{\robert}[1]{\textcolor{teal}{[Robert: #1]}}

\newcommand{\orest}[1]{\textcolor{orange}{[Orest: #1]}}

\newcommand{\lukasz}[1]{\textcolor{cyan}{[Łukasz: #1]}}

\definecolor{hlL}{rgb}{0.8 0.8 0.99}

\newcounter{highlight}

\newcounter{hlLR}

\newcounter{hlLIR}

\newcounter{hlLIIR}

\newcounter{Ahighlight}

\usepackage{scalerel,stackengine}
\stackMath
\newcommand\rwh[1]{%
\savestack{\tmpbox}{\stretchto{%
  \scaleto{%
        \scalerel*[\widthof{\ensuremath{#1}}]{\kern-.6pt\bigwedge\kern-.6pt}%
                  {\rule[-\textheight/2]{1ex}{\textheight}}
                              }{\textheight}%
}{0.5ex}}%
\stackon[1pt]{#1}{\tmpbox}%
}

\usepackage[hang,flushmargin]{footmisc}

\renewcommand{\epsilon}{\ensuremath\varepsilon}

\renewcommand{\phi}{\ensuremath{\varphi}}

\renewcommand{\maciej}[1]{}
\renewcommand{\lukasz}[1]{}
\renewcommand{\robert}[1]{}
\renewcommand{\orest}[1]{}

\ifconf
\fi

 \usepackage[normalem]{ulem}

\begin{document}

\title{\mytitle}


\ifnonbd

\ifconf

\author{Maciej Besta}\authornote{Corresponding Author}
\email{maciej.besta@inf.ethz.ch}
\affiliation{%
  \institution{ETH Zurich}
  \city{Zurich}
  \country{Switzerland}
}

\author{Łukasz Jarmocik}
\affiliation{%
  \institution{Cledar}
  \city{Wieliczka}
  \country{Poland}
}

\author{Orest Hrycyna}
\email{orest.hrycyna@ncbj.gov.pl}
\affiliation{%
  \institution{NCBJ Warszawa}
  \city{Otwock-Świerk}
  \country{Poland}
}

\author{Shachar Klaiman}
\affiliation{%
  \institution{BASF SE}
  \city{Ludwigshafen}
  \country{Germany}
}

\author{Konrad Maczka}
\affiliation{%
  \institution{Cledar}
  \city{Wieliczka}
  \country{Poland}
}

\author{Robert Gerstenberger}
\affiliation{%
  \institution{ETH Zurich}
  \city{Zurich}
  \country{Switzerland}
}

\author{J\"{u}rgen M\"{u}ller}
\affiliation{%
  \institution{BASF SE}
  \city{Ludwigshafen}
  \country{Germany}
}

\author{Piotr Nyczyk}
\affiliation{%
  \institution{IDEAS Research Institute}
  \city{Warsaw}
  \country{Poland}
}

\author{Hubert Niewiadomski}
\affiliation{%
  \institution{IDEAS Research Institute}
  \city{Warsaw}
  \country{Poland}
}

\author{Torsten Hoefler}
\affiliation{%
  \institution{ETH Zurich}
  \city{Zurich}
  \country{Switzerland}
}

\else

\author{Maciej Besta$^{1\dagger}$,
Łukasz Jarmocik$^{2,3}$,
Orest Hrycyna$^{2,3,4\dagger}$,
Shachar Klaiman$^{5}$,
Konrad Mączka$^{3}$,\\
Robert Gerstenberger$^{1}$,
Jürgen Müller$^{5}$,
Piotr Nyczyk$^{2,3}$,
Hubert Niewiadomski$^{2,3}$,
Torsten Hoefler$^{1}$}
\affiliation{\vspace{0.35em}$^1$ETH Zurich\quad
$^2$IDEAS Research Institute\quad
$^3$Cledar\quad
$^4$NCBJ Warszawa\quad
$^5$BASF SE \\
\country{{$^\dagger$}Corresponding authors}}
\fi



\renewcommand{\shortauthors}{Besta et al.}

\fi 

\begin{abstract}
Graphs are foundational across domains but remain hard to use without deep expertise. LLMs promise accessible natural language (NL) graph analytics, yet they fail to process industry-scale property graphs effectively and efficiently: such datasets are large, highly heterogeneous, structurally complex, and evolve dynamically. To address this, we devise a novel abstraction for complex multi-query analytics over such graphs. Its key idea is to replace brittle generation of graph queries directly from NL with planning over a \textit{Semantic Catalog} that describes both the graph schema and the graph operations. Concretely, this induces a clean separation between a \emph{Semantic Plane} for LLM planning and broader reasoning, and an \emph{Execution Plane} for deterministic, database-grade query execution over the full dataset and tool implementations. This design yields substantial gains in both token efficiency and task effectiveness even with small-context LLMs. We use this abstraction as the basis of the first LLM-enhanced graph analytics framework called \schemename. \schemenameAS achieves substantially higher success rates (e.g., 86\% over enhanced LangChain) and points toward the next generation of affordable and accessible graph analytics that unify LLM reasoning with database-grade execution over large and complex property graphs.
%
%
\end{abstract}


\maketitle

\section{INTRODUCTION}
\label{sec:intro}

\begin{figure*}[t]
\vspaceSQ{-0.5em}
 \centering
 \includegraphics[width=1.0\textwidth]{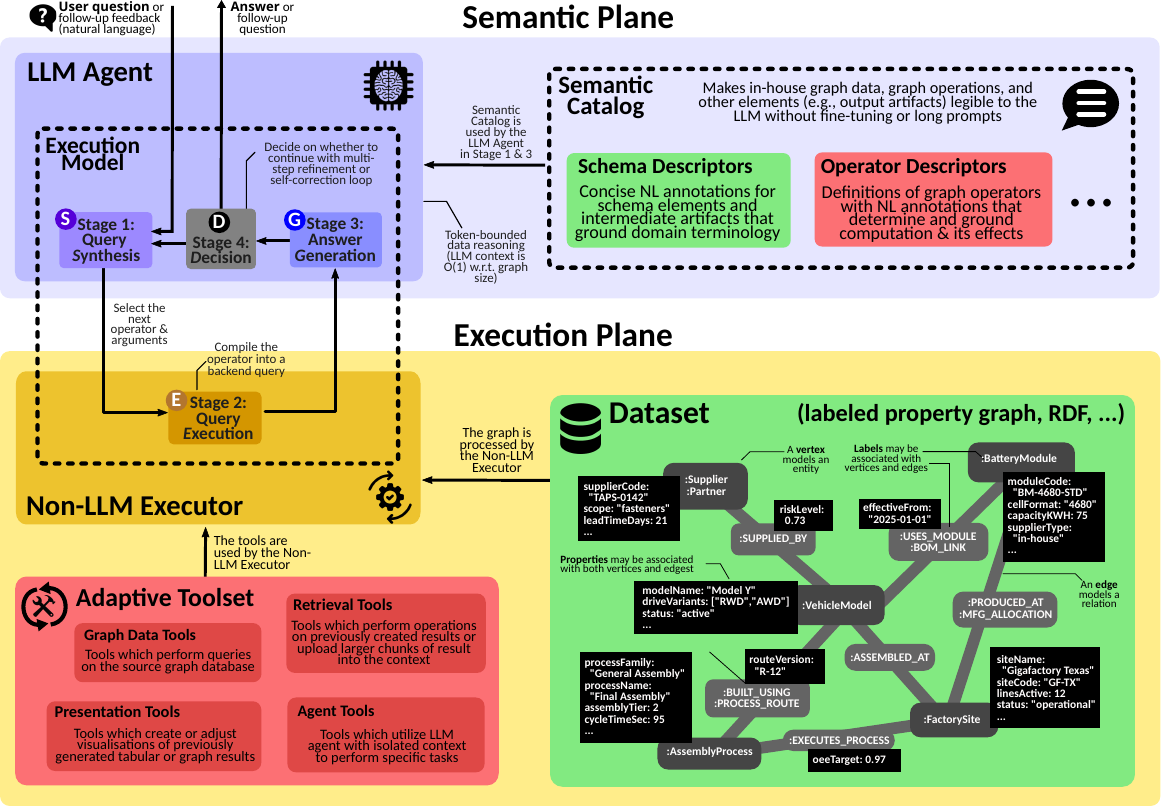}
 \vspace{-2.0em}
\caption{Overview of the key idea and architectural pillars behind \schemenameA: \textbf{Semantic Catalog} makes in‐house graph data, graph operations, and
other elements legible to the LLM without fine‐tuning or long prompts; separation of \textbf{Semantic} and \textbf{Execution Planes} allows the LLM to focus on high-level analytic reasoning and enables both effective and efficient data reasoning.}
\label{fig:idea}
\vspaceSQ{-0.5em}
\end{figure*}

\iftr
Graphs are critically important in academia and industry in numerous domains, including bioinformatics (e.g., analyzing protein structures), social sciences (e.g., studying human interactions), chemistry (e.g., designing chemical compounds), medicine (e.g., discovering drugs), cybersecurity (e.g., identifying intruders), healthcare (e.g., exposing fraudulent claims), web graph analysis (e.g., fostering accurate search services), and many others~\cite{besta2023demystifying, lumsdaine2007challenges, liu2025graph}. For example, graph databases are more powerful for representing and querying complex relationships than relational and document databases~\cite{angles2018introduction, davoudian2018survey, han2011survey, gajendran2012survey, gdb_survey_paper_Kaliyar}. Graph query languages like Cypher~\cite{francis2018cypher} and Gremlin~\cite{rodriguez2015gremlin} as well as graph databases and frameworks such as K{\`u}zu~\cite{feng2023kuzu} and GDI~\cite{besta2023graph} enable effective pattern matching~\cite{Angles2012, Robinson2015}, with linear time traversal performance compared to costly relational joins, fostering practical solutions~\cite{Robinson2015}. Beyond graph databases, graph analytic frameworks and schemes such as GraphMineSuite~\cite{gms} or ProbGraph~\cite{besta2022probgraph} have also demonstrated their ability to capture patterns and make predictions that would be impossible or less accurate using traditional flat data analytics~\cite{satish2014navigating, sakr2021future, besta2019practice}. Yet, \textit{graphs impose high learning curve} as they usually demand skill in graph theory, graph query languages, and advanced modeling decisions.

Large Language Models (LLMs) offer a promising opportunity to make graph analytics more accessible by enabling interaction through natural language, without requiring deep technical expertise in graph theory or query languages~\cite{zhao2023survey, minaee2024large, besta2025reasoning}. However, realizing this vision at industry scale remains challenging. Most LLMs are trained primarily on natural language (NL) and code and thus struggle with precise graph pattern matching, schema-aware query formulation, and multi-hop traversal. This gap is amplified in real-world settings, where industrial graphs are highly heterogeneous and governed by evolving schemas. A seemingly straightforward strategy based on injecting graph structure into the prompt is infeasible as production graphs can span millions of nodes and edges. Practical systems must also support database-grade operations including filtering, aggregation, temporal constraints, transactional updates, and structured result presentation.  \textit{The grand challenge is to enable NL-driven graph analytics that is both effective and efficient under tight token and latency budgets.}
\fi

We address this challenge by introducing a \textit{graph processing abstraction} for large and complex heterogeneous property graphs (\textbf{contribution~1}). Its key component is a \textbf{Semantic Catalog}~\includegraphics[scale=0.2,trim=0 16 0 0]{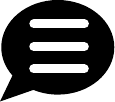}, a compact semantic interface that supports the \emph{Semantic Plane}, where an LLM agent plans and interprets analytics requests (top of Figure~\ref{fig:idea}). The catalog is motivated by a practical observation: directly translating NL into backend graph queries is brittle and often misaligned with complex graph schemas and graph access patterns. It makes both the \emph{graph data} and the \emph{graph operations} legible to an LLM {without fine-tuning} or long-context prompting. It contains (a) \emph{Schema Descriptors}, concise NL annotations for schema elements that ground domain terminology and reduce ambiguity even for specialized in-house schemas, and (b) \emph{Operator Descriptors}, a list of intermediate graph operators (e.g., $k$-hop traversal, temporal slicing, neighborhood aggregation) with explicit NL annotations that clarify operator behavior with respect to the graph data. Together, these components replace brittle generation of graph queries from NL with robust planning over a Semantic Catalog.

\enlargethispage{\baselineskip}

To turn catalog-grounded plans into efficient graph analytics, we also require an explicit \textit{execution model} that bridges Semantic and Execution Planes (Figure~\ref{fig:idea}, bottom). For this, we harness TAG (Table-Augmented Generation)~\cite{biswal2024text2sql}, a recent execution model for relational databases augmented with LLMs, which we extend to graph processing and multi-step queries (\textbf{contribution~2}). TAG decomposes NL questions into \textbf{\emph{\textcolor{violet}{query synthesis}}}~\includegraphics[scale=0.2,trim=0 16 0 0]{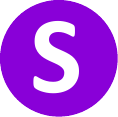}, \textbf{\emph{\textcolor{orange}{query execution}}}~\includegraphics[scale=0.2,trim=0 16 0 0]{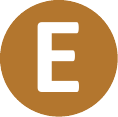}, and \textbf{\emph{\textcolor{blue}{answer generation}}}~\includegraphics[scale=0.2,trim=0 16 0 0]{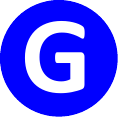}. In our instantiation, {synthesis} and {generation} operate in the Semantic Plane: {synthesis} maps NL to typed operator invocations drawn from the Semantic Catalog, and {generation} produces answers from compact intermediate views. {Execution} operates in the Execution Plane: a Non-LLM runtime deterministically compiles and runs these invocations on the graph backend (minimizing token costs). We further add a \textbf{\emph{Decision}} stage~\includegraphics[scale=0.2,trim=0 16 0 0]{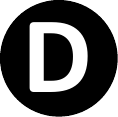} to make the loop multi-step, enabling token-bounded refinement and self-correction: intermediate results are materialized out-of-context and referenced across steps, while only compact $O(1)$ summaries and selected subsets are returned in-context.

We implement this abstraction in \textbf{\schemenameA}, the first LLM-enhanced system for interactive processing of production-scale property graphs, and we detail its design (\textbf{contribution~3}). Using \schemenameA, we develop and detail five real-world industry use cases drawn from an electric vehicle (EV) manufacturing setting, illustrating the end-to-end workflow and the kinds of multi-query analytics enabled by our Semantic/Execution Plane design (\textbf{contribution~4}). We compare \schemenameAS against established LLM-based graph analytics baselines (\textbf{contribution~5}). Across diverse workloads, \schemenameAS consistently achieves higher success rates (86\% over an enhanced LangChain~\cite{Langchain}) while maintaining low token costs and runtimes. Ablation studies isolate the effect of the Semantic Catalog, showing substantial gains in reliability and query correctness. Overall, \schemenameAS demonstrates how to unify LLM reasoning with database-grade execution for graph analytics, and provides an operational blueprint for multi-hop TAG-style systems over large-scale heterogeneous property graphs.

\section{LANDSCAPE OF GRAPHS \& LLMS}
\label{sec:landscape}

No existing graph-oriented scheme enables interactive graph analytics over industry-scale graphs without demanding expertise in graph query/theory. Existing approaches either assume small, in-context graphs, focus on predictive representation learning rather than symbolic query execution, or are built for non-graph backends, and thus fundamentally cannot support precise, multi-hop, and schema-aware queries.

First, several works have explored the {intersection of graphs and LLMs}, but focused on simple prompting or fine-tuning to perform basic algorithmic tasks (e.g., graph connectivity) or learning tasks (e.g., node classification) on small graphs that fit entirely in the model's context~\cite{peng2024chatgraph, li2024graphlearning, li2024preliminary, li2024graphintelligence, bonifati2025roadmap, fan2024graph, chen2024exploring, shang2024survey}. Some studies have also addressed the generation of graph queries in languages such as SPARQL~\cite{sparql-ref}, emphasizing the syntactic and semantic correctness of query generation~\cite{rangel2024sparql, brei2024leveraging, mountantonakis2025generating, meyer2024assessing, arazzi2025augmented, schwabe2025q, li2024linkq, kakalis2024advancing}. However, none of these schemes considers the broader scope of industry-scale graph analytics. These approaches assume that graph structure or large schema fragments can be serialized into an LLM context and that the model can emit correct free-form queries; yet, they break down on large graphs due to context limits, inability to build and use tools, or to handle changing datasets.


\iftr
Beyond these directions, in the broader domain of {graph representation learning (GRL)}, substantial efforts have been devoted to the development of {Graph Foundation Models (GFMs)}~\cite{zhang2023graph, zhao2024graphany, chen2024text, fu2024parametric, zhu2025graphclip, xia2024anygraph, galkin2023towards, liu2025graph, mao2024position, xia2024opengraph} and {Graph Neural Networks (GNNs)}~\cite{wu2020comprehensive, zhou2020graph, zhang2020deep, besta2022motif, besta2022neural, besta2024hot, besta2023high, besta2025demystifyinghigher, besta2023parallel, chami2020machine, hamilton2017representation, bronstein2017geometric, kipf2016semi, xu2018powerful, wu2019simplifying, bazinska2023cached}, 
\fi
including important subclasses such as Graph Transformers~\cite{shehzad2024graph} and Graph Attention~\cite{velickovic2018graph}, as well as other {graph prompting} schemes such as GraphPrompt~\cite{liu2023graphprompt}.
However, these designs are built for predictive tasks rather than symbolic querying, so they cannot natively enumerate result sets, perform constraint- and time-aware joins, or explain intermediate reasoning steps. Their stochastic, embedding-based inference is ill-suited to interactive, exact analytics over large-scale graphs without exposing users to low-level graph modeling and training details.


\begin{figure}[t]
 \centering
 \iftr
 \includegraphics[width=0.48\textwidth]{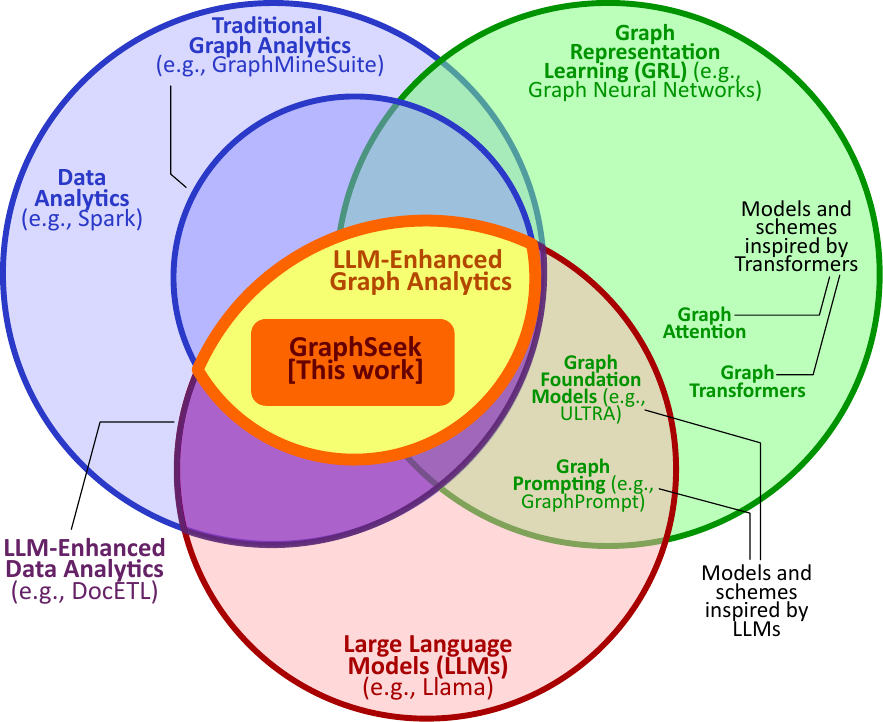}
 \fi
 \vspaceSQ{-1.0em}
\caption{\textbf{Overview of research areas related to this work.} \schemenameAS sits on the overlap of \textcolor{blue}{Traditional Graph Analytics}, \textcolor{red}{Large Language Models}, and \textcolor{green}{Graph Representation Learning}.}
\label{fig:landscape}
\vspaceSQ{-2.0em}
\end{figure}

\iftr
Finally, in data analytics beyond graphs, there has been significant growth in AI frameworks~\cite{weng2024insightlens, lee2025semantic, chang2025approximating, zeighami2025llm, biswal2024text2sql, chen2023data, qiu2024tqa, dorbani2025beyond}. 
\fi
\ifconf
\else
This includes document analysis (e.g., Aryn~\cite{anderson2024design}, DocETL~\cite{shankar2024docetl}, Palimpzest~\cite{liu2025palimpzest}, PalimpChat~\cite{liu2025palimpchat}), tabular data (e.g., InsightPilot~\cite{ding2023insightpilot}, CoddLLM~\cite{zhang2025coddllm}, Pneuma~\cite{balaka2025pneuma}, Chat2data~\cite{zhao2024chat2data}, LOTUS~\cite{patel2024semanticoperators}, DB-GPT~\cite{xue2024demonstration}), video processing~\cite{10.14778/3685800.3685916}, and others~\cite{devunuri2024transitgpt}.

\fi
These frameworks process unstructured or tabular data using RAG and agents, but lack property-graph traversal semantics (typed nodes/edges, k-hop paths, neighborhood aggregations, etc.) and a dedicated graph executor; scaling them to large graphs typically relies on long-context prompting or vector retrieval, which cannot guarantee precise, complete answers or interactive latency without reintroducing explicit graph-query expertise.


We overview all these domains in Figure~\ref{fig:landscape}.
\textit{We conclude that no existing work has yet proposed a comprehensive framework that leverages {generative AI capabilities} for {more effective and efficient} processing of {realistic industry-scale complex graph datasets}.}

\section{GRAPH PROCESSING ABSTRACTION}
\label{sec:execution-model}

\iftr
We introduce a graph processing abstraction that makes NL-based analytics over {property graphs} both robust and scalable. The abstraction has two tightly coupled components: (i) an \emph{execution model} that extends TAG~\cite{biswal2024text2sql} to \emph{multi-step} graph analytics with self-correction, and (ii) a \emph{Semantic Catalog} that serves as a compact interface between language and the graph backend.
The model retains TAG's three conceptual stages (query synthesis, query execution, and answer generation) but makes two changes essential for graph analytics: (1) synthesis produces \emph{typed operator invocations} rather than free-form backend queries, and (2) the pipeline is \emph{iterative}, producing and consuming intermediate results across multiple steps.
\fi
Together, they replace brittle direct graph-query generation from NL with catalog-grounded planning in the Semantic Plane, while delegating all query execution to the Execution Plane with exact database semantics.

\subsection{Execution Model}
\label{subsec:exec-model}

We model the data source as a \textbf{labeled property graph (LPG)} \(G = (V, E, \lambda_V, \lambda_E, \pi)\), where \(V\) and \(E\) are nodes and edges, \(\lambda_V(v)\) and \(\lambda_E(e)\) map to (possibly multiple) labels/types, and \(\pi\) maps nodes/edges to key--value properties. The schema \(\mathcal{S}\) specifies allowed node/edge types, property keys and value domains, and relationship signatures. A graph backend exposes an executable query language (e.g., Cypher) and a transactional runtime.

Let \(\mathcal{L}\) be a library of \textbf{typed semantic operators}. Each operator \(l \in \mathcal{L}\) has a compilation procedure into one or more backend queries. Operators are \emph{not} backend queries themselves; they form an intermediate representation designed for controllable synthesis.
Operators in the catalog fall into four groups: \emph{graph data} operators that issue queries against the source graph database and materialize results; \emph{retrieval} operators that operate over previously materialized results and, when needed, upload larger result chunks into the LLM context; \emph{presentation} operators that create or refine visualizations of prior tabular or graph outputs; and \emph{agent} operators that invoke an LLM agent with an isolated context to perform specific sub-tasks.


Given a natural language request \(R\) over \(G\), the system maintains a step-indexed interaction state
\iftr
\[
\sigma_t \;=\; (R, G, \mathcal{S}, \mathcal{L}, \mathcal{C}, \mathcal{A}_t, \mathcal{T}_t),
\]
\fi
where $t$ is the current step, \(\mathcal{C}\) is the Semantic Catalog, \(\mathcal{A}_t\) is an artifact store of executor-produced intermediate results represented by \emph{handles} (node/edge sets, projected tables, aggregate summaries, subgraphs), and \(\mathcal{T}_t\) is a compact execution trace (messages, operator calls, and outcomes) up to step \(t\).
This state makes explicit the multi-step contribution: the system composes dependent operator calls across steps and can revise earlier choices based on execution feedback.

Each step \(t\) decomposes into four stages:
\vspaceSQ{-1em}

\small
\begin{align}
\textbf{Synthesis (Stage 1) \includegraphics[scale=0.2,trim=0 16 0 0]{S.pdf}:} \quad & \textsf{syn}_t(\sigma_t;\ \mathcal{C}) \rightarrow a_t \\
\textbf{Execution (Stage 2) \includegraphics[scale=0.2,trim=0 16 0 0]{E.pdf}:} \quad & \textsf{exec}(a_t, G) \rightarrow r_t \\
\textbf{Generation (Stage 3) \includegraphics[scale=0.2,trim=0 16 0 0]{G.pdf}:} \quad & \textsf{gen}_t(R,\ \mathcal{T}_t,\ r_t;\ \mathcal{C}) \rightarrow y_t \\
\textbf{Decision (Stage 4) \includegraphics[scale=0.2,trim=0 16 0 0]{D.pdf}:} \quad & \textsf{dec}_t(y_t,\ r_t,\ \mathcal{T}_t) \rightarrow \{\textsf{halt},\textsf{continue}\}
\end{align}
\normalsize
where \(a_t\) is a typed operator invocation \(a_t=\langle o_t,\theta_t\rangle\) with operator \(o_t\in\mathcal{L}\) and arguments \(\theta_t\), \(r_t\) is the executor response, and \(y_t\) is a candidate answer that may be final or may indicate the need for additional evidence (e.g., ``need another traversal to resolve ambiguity'').
Crucially, \textsf{syn} and \textsf{gen} are \emph{catalog conditioned}: in synthesis, the LLM selects and instantiates operators, and interprets returned artifacts, relying on Schema and Operator Descriptors.

The Non-LLM Executor deterministically compiles \(a_t\) into one or more backend queries (e.g., Cypher), executes them over \(G\), and returns a full result object persisted out-of-context as a handle in \(\mathcal{A}_{t+1}\), and a compact in-context observation (small samples, aggregates, statistical summaries) optionally augmented with Artifact Descriptors from \(\mathcal{C}\). This design keeps the LLM context \(O(1)\) with respect to \(|V|+|E|\) while retaining exact database semantics.

The loop continues while \(\textsf{dec}_t\) returns \textsf{continue} and stops when it returns \textsf{halt} (emitting \(y_t\) as the final answer) or a step budget is reached. Decision enables token-bounded self-correction: if \(r_t\) indicates an issue (incorrect operator choice, empty results, etc.) then \(\textsf{dec}_t\) selects \textsf{continue}, and \textsf{syn} revises \(a_{t+1}\) based on \(\mathcal{S}\), \(\mathcal{C}\), and the trace \(\mathcal{T}_t\); otherwise it returns the answer.

\subsection{Semantic Catalog}
\label{subsec:catalog}

The Semantic Catalog \(\mathcal{C}\) is a compact, queryable interface that grounds both \emph{what the data means} and \emph{what computations mean}. We use the term \textit{Semantic Descriptor} to denote a short NL annotation attached to an object, and distinguish two main descriptor families: \textbf{Schema Descriptors} and \textbf{Operator Descriptors}. Both are consumed by the execution model: \textsf{syn} uses Schema/Operator Descriptors to map \(R\) to \(a_t\), and \textsf{gen} uses Schema/Artifact Descriptors to verbalize \(r_t\) and to decide whether to continue.

\iftr
While not required for correctness, the Semantic Catalog improves disambiguation of symbolic structures, aligns user intent with schema elements, and supports interpretation of complex outputs. We use it to augment standard property-graph schemas into \emph{semantic} graph schemas, enriching node types, relationship types, and attributes with domain-specific natural language descriptors that capture intent and context. These descriptors are authored by experts and/or refined by LLMs, and encode proprietary, company-specific semantics that no general model can reliably infer \emph{a priori}. In our setting, their footprint is small (about $2\mathrm{k}$ tokens for the entire graph). \schemenameAS exposes an API to inject such knowledge (e.g., business glossaries, process labels, codebooks). To reduce authoring cost and variability, we use a hybrid pipeline: drafts are produced from schema introspection (types, constraints), sample values, existing documentation, and prior query logs; LLMs propose candidates under strict schema-grounded templates, and humans validate them before integration.
\fi

\begin{figure*}[t]
 \centering
 \includegraphics[width=1.0\textwidth]{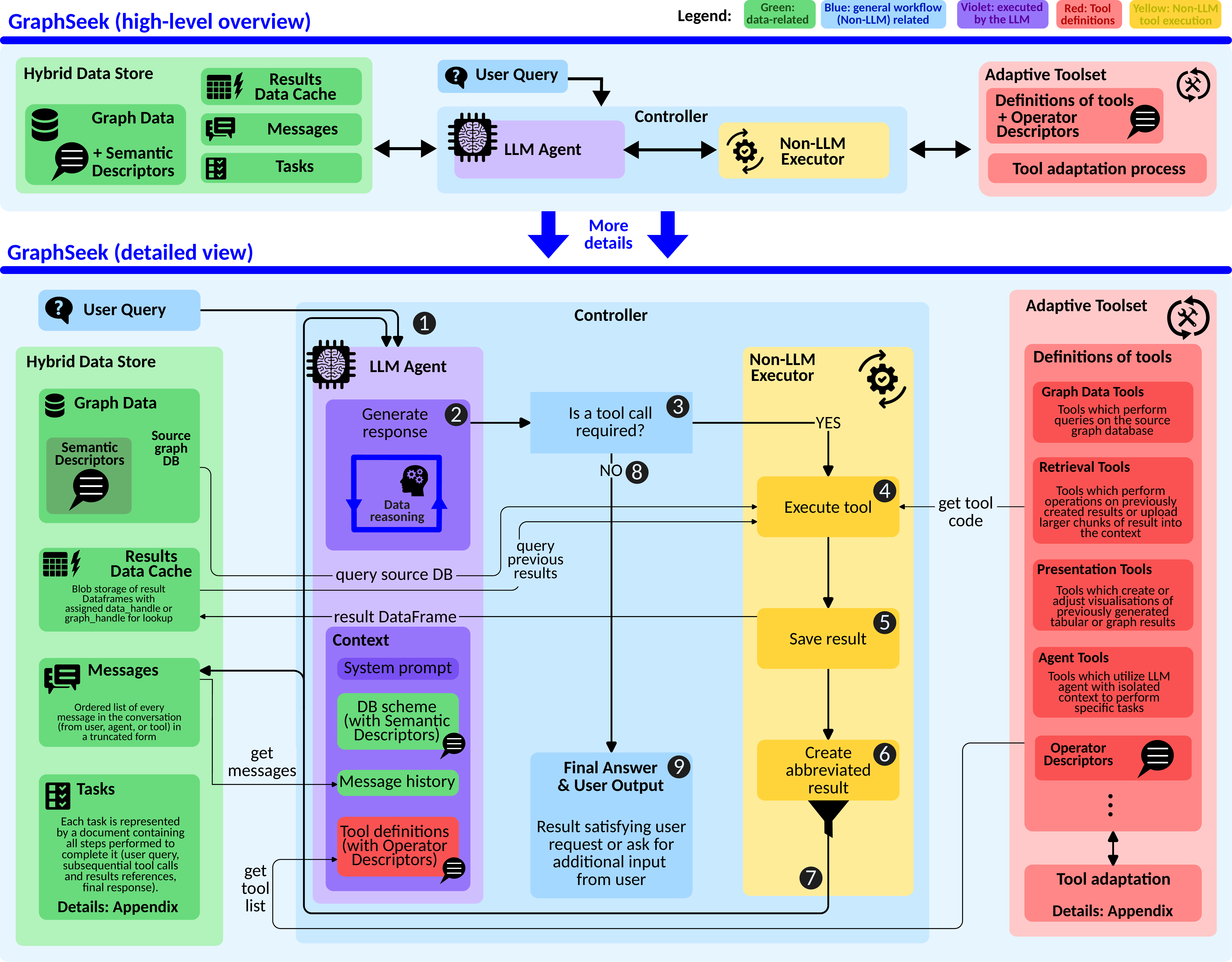}
  \vspace{-2em}
\caption{\textbf{Overview of the \schemenameS framework; detailed workflows for the task lifecycle and tool adaptation are in Appendix~\ref{sec:app:arch}.}}
\label{fig:overview}
\vspaceSQ{-1em}
\end{figure*}

\section{GRAPHSEEK: DESIGN \& ARCHITECTURE}
\label{sec:arch}

We now overview \schemenameAS (Figure~\ref{fig:overview}) that implements our abstraction. \schemenameAS comprises three cooperating modules: the \textbf{Controller}, the \textbf{Hybrid Data Store}, and the \textbf{Adaptive Toolset}. The {Controller} contains an LLM Agent (Semantic Plane) and a Non-LLM Executor (Execution Plane). The agent interprets user intent and instantiates {operator invocations} from the Semantic Catalog; these invocations are realized via \emph{tools} from the Adaptive Toolset and are compiled and executed against the graph backend by the Non-LLM Executor. Results flow into the Hybrid Data Store. The Controller can also propose edits to tool definitions, which are implemented after validation and optional human approval.

\subsection{\mbox{\hspace{-0.5em}Controller \& Token-Efficient Data Reasoning}}
\label{sec:data-reasoning}

To minimize token costs \& latency and maximize effectiveness, the LLM Agent never operates on raw graph content nor emits free-form Cypher. Instead, given a user request $x$ and the Semantic Catalog $\mathcal{C}$, it produces a \emph{typed operator call} $l(\theta)$, where each $l \in \mathcal{L}$ is an operator annotated in $\mathcal{C}$ and $\theta$ are schema-validated arguments. Operators are \emph{implemented by tools}: each operator call $l(\theta)$ maps to a concrete tool invocation $t=\tau(l,\theta)$ in the Adaptive Toolset, whose full code lives in the Execution Plane.
The Non-LLM Executor deterministically compiles each tool invocation into backend queries (e.g., Cypher), executes them, post-processes results, and returns only compact $O(1)$ summaries $s_t=\sigma(r_t)$ to the agent for the next step $t$. Full payloads $r_t$ are stored out-of-context and retrieved only on demand (e.g., via retrieval tools) when additional detail is required. This architecture separates the {LLM reasoning in the Semantic Plane} and the Non-LLM {execution in the Execution Plane}.
We present an example compact graph summary in Appendix~\ref{sec:app:arch}.


\subsection{Hybrid Data Store}

The Hybrid Data Store provides persistent state across the two planes. First, a graph database stores the input graph $G$. Second, all tool outputs are written to a \textbf{Results Data Cache}. Only compact summaries $s=\sigma(r)$ are returned in-context; full payloads $r$ (tables, graph fragments, files) can be retrieved on demand.
Summaries may include tabular outputs (schema/column metadata, head/tail previews, etc.), textual outputs (truncated bodies with key metadata), and graph outputs (node/edge counts, small visualizable subgraphs, etc.).
Third, a \textbf{message history} preserves conversational context for consistent multi-step reasoning. Finally, a \textbf{task store} records task definitions, execution traces, and logs for inspection and reuse. Each task trace captures the chain
$
(x \rightarrow \pi \rightarrow \{t_t\} \rightarrow \{r_t\} \rightarrow \{s_t\} \rightarrow y),
$
enabling end-to-end visualization and fine-grained debugging/auditing. Traces and conversations are interactive artifacts: users can revisit any step to branch, refine, or replay segments.


\subsection{Adaptive Toolset}

The Adaptive Toolset supplies the Execution Plane with the concrete \emph{tool implementations} that realize Catalog Operators. Tools are schema-validated, self-documenting function specifications that the LLM instantiates (via operator calls) and the Non-LLM Executor deterministically compiles into precise backend actions.
\iftr
The library covers routines for {graph access} (entity lookups, neighborhood and multi-hop traversals, aggregations, temporal slicing, etc.), {result retrieval/manipulation} (loading full outputs or slices, conversions between graph and tables, constrained dataframe operations, joins/merges, etc.), {presentation} (tables, charts, etc.), and {agent utilities} (isolated-context subtasks, controlled fallbacks).
\fi
The toolset is extensible. Critically, it is also self-adaptive: at inference time the LLM may propose edits to a tool; the system validates candidates against contracts (e.g., JSON Schema), compiles and dry-runs against the live schema, versions the candidate, and (optionally) gates promotion via human approval.

\subsection{Details of System Workflow}
\label{sec:workflow}

The execution of analytics tasks in \schemenameAS is detailed in Figure~\ref{fig:overview} \iftr(center). \fi It begins when the user submits a natural language request $x$ via the chat interface~\includegraphics[scale=0.2,trim=0 16 0 0]{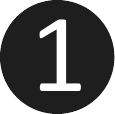}. In the Semantic Plane, the LLM Agent processes $x$ under a contextual prompt composed of the system instruction, the Semantic Catalog $\mathcal{C}$, prior interaction history, and a structured list of callable tools.

The LLM responds with either a plain-text message $y$ or a structured operator invocation (mapped to a tool call)~\includegraphics[scale=0.2,trim=0 16 0 0]{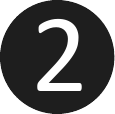}. In the latter case~\includegraphics[scale=0.2,trim=0 16 0 0]{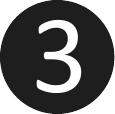}, the system forwards the call to the {Non-LLM Executor} (Execution Plane), which executes the specified tool with arguments~\includegraphics[scale=0.2,trim=0 16 0 0]{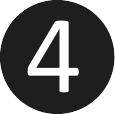}. Tool logic may query the source graph database, operate over previously materialized intermediate results referenced by IDs, or run custom computation logic.

\iftr
The output of tool execution~\includegraphics[scale=0.2,trim=0 16 0 0]{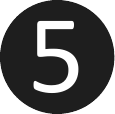} (typically a graph fragment or structured table) is stored in the {Result Data Cache} as one or more objects (e.g., \textsf{DataFrames}, graph views). To enable coherent multi-step reasoning within token limits, a compact natural language summary $s=\sigma(r)$ is generated~\includegraphics[scale=0.2,trim=0 16 0 0]{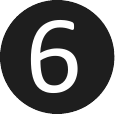} and appended to the message history passed to the LLM Agent in the next turn~\includegraphics[scale=0.2,trim=0 16 0 0]{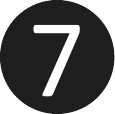}. The loop continues until the LLM Agent emits a plain-text response $y$~\includegraphics[scale=0.2,trim=0 16 0 0]{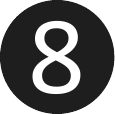}, presented to the user along with relevant visualizations (tables, charts, graphs)~\includegraphics[scale=0.2,trim=0 16 0 0]{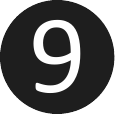}.
\fi

\section{GRAPHSEEK: EXAMPLE USE CASES}
\label{sec:usecases}

\iftr
We provide five use cases of \schemenameAS and also pinpoint \schemenameA's advantages over traditional approaches (due to space constraints, detailed listings of each use case are in Appendix~\ref{sec:app:use-cases}).
We select these use cases to showcase the ability of \schemenameAS to automatically detect and execute graph analytics patterns such as {aggregation}, {traversal}, graph-level comparison, and fuzzy property lookups. Moreover, they highlight the capability of the system to conduct multi-step reasoning by splitting a complex question into sub-queries, and to accelerate reasoning by identifying parallelization opportunities in tool execution.
\fi

\subsection{A Running Example: EV Manufacturing}
\label{sec:graph-example}

\iftr
The use cases harness an anonymized but representative \emph{electric vehicle (EV) manufacturing graph}\footnote{To ensure anonymity of the in-house datasets, we fully change the naming and in-house data specifics from the original one into EV manufacturing; however, we keep the original general graph structure and make sure that all the insights and observations remain valid.}, see Figure~\ref{fig:idea} (bottom). Production is organized into sequential assembly tiers (tier~0 corresponds to a finished vehicle model; higher tiers represent upstream subassemblies and modules). At each tier, an \emph{assembly process} executed on a specific \emph{assembly line} within a \emph{factory site} transforms \emph{components} (e.g., battery modules and prior-tier drive assemblies) into a new \emph{subsystem}, potentially generating \emph{auxiliary modules} or scrap; intermediate subsystems may be transferred across factory sites within a single vehicle's manufacturing flow. The graph links vehicle models, intermediate assemblies, component modules, assembly lines, factory sites, organizational units, and regions, together with temporal annotations (e.g., month/quarter, plan vs.\ actual) and cost/quantity information.
\fi

The schema captures key entities and relationships involved in the design, production, and deployment of modular EV components. It defines node types such as \textsf{BatteryModule} (individual energy storage units), \textsf{DriveAssembly} (motor, inverter, and transmission components), \textsf{VehicleModel} (complete EV configurations), \textsf{FactorySite} (physical production facilities), and \textsf{AssemblyLine} (production lines used to build specific component types). Relationships include \textsf{INTEGRATED\_IN}, \textsf{OUTPUTS}, \textsf{INSTALLED\_AT}, and \textsf{CONNECTED\_TO}. Example relation types could be \textsf{(:DriveAssembly)} \textsf{-} \textsf{[:INTEGRATED\_IN]} \textsf{->} \textsf{(:VehicleModel)} or \textsf{(:AssemblyLine)} \textsf{-} \textsf{[:INSTALLED\_AT]} \textsf{->} \textsf{(:FactorySite)}. 

Each node type includes domain-specific attributes that enable detailed analytics: e.g., \textsf{BatteryModule} nodes may store \textsf{energyDensity} and \textsf{configVersion}; \textsf{DriveAssembly} nodes may include \textsf{efficiencyRate} or \textsf{assemblyCost}; and \textsf{FactorySite} nodes may contain operational metadata such as \textsf{siteCode} or region.

\subsection{Example Semantic Descriptors}

In the EV graph, for instance, the \texttt{DriveAssembly} node may be annotated as: \textit{``Represents a partially assembled propulsion subsystem within a multi-stage vehicle production process; it may consume earlier components (e.g., electric motors) and supply integrated assemblies to final vehicle models.''} Likewise, the \texttt{INTEGRATED\_IN} relation might be described as: \textit{``Denotes that a component or subsystem, such as a battery or drive assembly, is physically and functionally incorporated into a higher-level product during assembly.''} This semantic overlay allows the LLM to distinguish functionally distinct but structurally similar elements (such as battery packs and drive systems) not merely by label but by their role in the assembly flow.

\subsection{Example Use Cases}

\iftr
We now illustrate \schemenameAS on five real-world EV manufacturing requests. Each use case is written as an \emph{industry question}, but is also representative of (i) a recurring \emph{graph access pattern} (e.g., filtered selection, neighborhood traversal, aggregation), and (ii) a related \emph{data-reasoning feature} of our Semantic/Execution Plane design (e.g., catalog grounding, multi-step reasoning, parallel tool calls, self-correction). In all cases, the Semantic Plane compiles the request into a short plan $\pi=(o_1(\theta_1),\ldots,o_T(\theta_T))$ of typed operator invocations from the Semantic Catalog, while the Execution Plane deterministically executes corresponding tools on the graph backend and returns compact summaries (with full results cached out-of-context).
\fi

\subsubsection{\textbf{Aggregating Upstream Modules per Factory.}}

The first example shows answering a question \textit{``How many baseline drive assemblies do we produce per factory?''} (Listing~\ref{lst:e1}). This question requires \textbf{constructing an aggregate view over the graph}. The LLM agent correctly interprets {``baseline assemblies''} as drive assemblies at the initial process tier (\textsf{processTier = 0}), selects an appropriate graph tool, and returns a human-readable summary.

This process stresses the pattern of attribute-filtered node selection followed by group-by aggregation. The data reasoning pattern is \textbf{schema/terminology grounding} via the Semantic Catalog: the agent resolves {base-tier/baseline} to a typed predicate ($\textsf{assemblyTier}=0$), selects the correct node type (\textsf{DriveAssembly}) and grouping key (\textsf{factoryCode}), and emits a typed operator invocation consistent with the catalog signature.

In \schemenameA, the user provides only a high-level, domain oriented NL query. Without \schemenameA, one would need to understand the exact graph schema and write a Cypher query like the one in Listing~\ref{lst:e1}. Such queries must also adapt to differences in semantics or structure across domains. For non-experts, learning schema, properties, and syntax is a significant barrier. In \schemenameA, the LLM handles this mapping and query generation automatically, while the graph engine performs efficient execution.

\subsubsection{\textbf{Inspecting a Production Plan for a Vehicle.}}

\iftr
Here, a user requests a \textit{``unique production plan''} for a given vehicle model (Listing~\ref{lst:e2}). The system retrieves, composes, and explains a full subgraph representing the plan. The user expresses a high-level question about the plan for model \textsf{PX-206}, and the system resolves the name to a \textsf{VehicleModel} node, {traverses its neighborhood} (e.g., \textsf{BatteryModule}, \textsf{DriveAssembly}, \textsf{AssemblyLine}, \textsf{FactorySite}), and returns both a human-readable graph (via \textsf{graph\_handle}) and a textual explanation.
\fi

The stressed data-reasoning feature is \textbf{compositional operator planning} plus \textbf{result interpretation}: the Semantic Plane resolves the model name, selects a plan-extraction operator, and then explains the returned graph handle using the Schema/Operator Descriptors. The Execution Plane performs the \textbf{neighborhood traversal} and returns a compact structural summary together with a \textsf{graph\_handle} that the UI can render.

A manual query requires understanding which labels and relationships compose a \textit{production plan} (e.g., \textsf{CONSUMED\_IN}, \textsf{GENERATES}, \textsf{PROCESSED\_AT}), and encoding those paths. This often involves iterative refinement to capture cost types, uniqueness constraints, and correct traversal depth. \schemenameAS handles all of this with the LLM’s operator selection and executor's compilation.

\subsubsection{\textbf{Identifying a Factory under Cost Criteria.}}

Here, the user asks: \textit{``Which factory produces the module with the highest market price?''} (Listing~\ref{lst:e3}). \schemenameAS decomposes this into three steps: compute the maximum market price, find the module with that price, and determine its production site. When an equality filter fails due to floating-point mismatch, the system automatically corrects itself by switching to an ordering-based query.

This example illustrates \textbf{multi-step reasoning} combined with \textbf{automatic self-recovery from errors}, as well as type reasoning and chaining of dependent tool calls. A human analyst would need to carefully manage numeric precision and manual joins; \schemenameAS orchestrates all this seamlessly.

\subsubsection{\textbf{Comparison \& Synthesis of Multiple Production Plans.}}

Next, the request is \textit{``Check unique production plans for three vehicle models whose names start with `B', and tell me which one uses the largest number of modules.''} (Listing~\ref{lst:e4}). Here, \schemenameAS can issue \textbf{multiple tool calls concurrently} and compare their outputs. The system performs fuzzy prefix matching, generates three separate calls to retrieve production plans, computes the number of distinct inputs per plan, and returns a synthesized comparison.

Without \schemenameA, such reasoning needs repeated queries, explicit name resolution, and manual aggregation across results. With tool modularity and executor-driven graph traversal, the LLM handles this as a single orchestration problem.

\subsubsection{\textbf{Self-Correction \& Human-In-The-Loop.}}

Finally, the request is a two-turn interaction: \textit{``What is the highest module cost?''} followed by \textit{``Which module has this highest cost?''} (Listing~\ref{lst:e5}). The user first requests the highest module cost, and then asks which module has that cost. After obtaining $\max(\textsf{unitCost})$, the agent initially tries an equality filter at rounded precision, receives an empty result, and then switches to a ranking-based operator (\mbox{top-$1$} by \textsf{unitCost}) that is stable under floating-point representation. The executor deterministically runs the revised query, and the agent returns the correct module along with relevant attributes (tier, key), maintaining a clean semantic interaction while the underlying strategy adapts automatically.
This use case showcases \textbf{human-in-the-loop} combined with \textbf{automatic error recovery}.

In traditional systems, such errors could confuse non-experts or require post-processing. \schemenameAS shifts strategy automatically and shields the user from underlying data processing details. Simultaneously, it enables interactive analytics, so that the user can follow-up on requests, facilitating novel insights into the data.

\section{EVALUATION}
\label{sec:eval}

\begin{figure*}[t]
\vspaceSQ{-0.5em}
  \centering
  \includegraphics[width=1.0\textwidth]{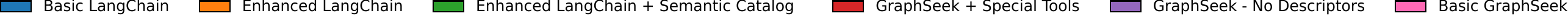}
  \begin{subfigure}[t]{\textwidth}
  \includegraphics[width=1.0\textwidth]{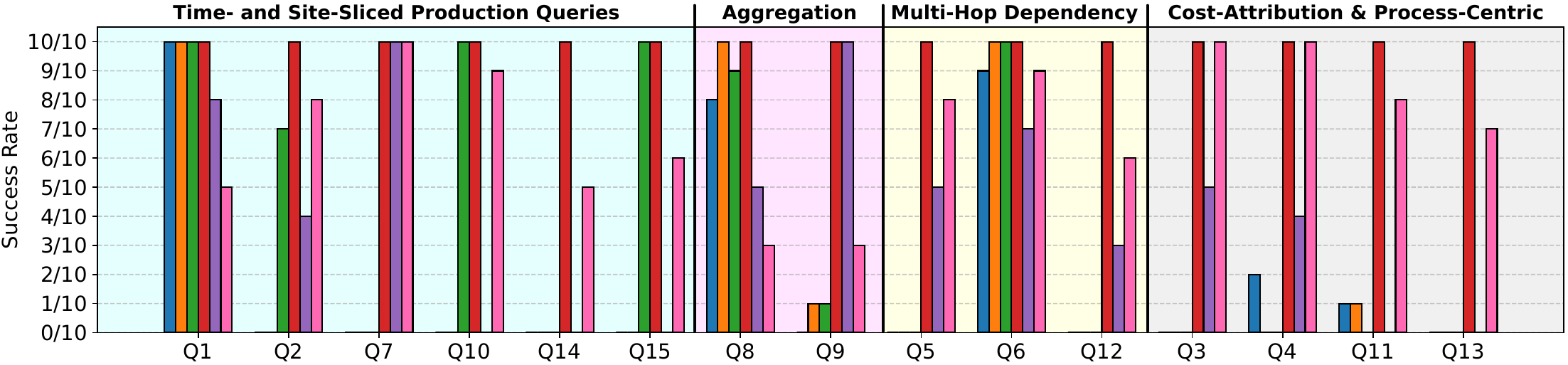}
  \vspace{-1.5em}
  \caption{Success rate.}
  \label{fig:eval-success}
  \end{subfigure}
  \begin{subfigure}[t]{\textwidth}
  \includegraphics[width=1.0\textwidth]{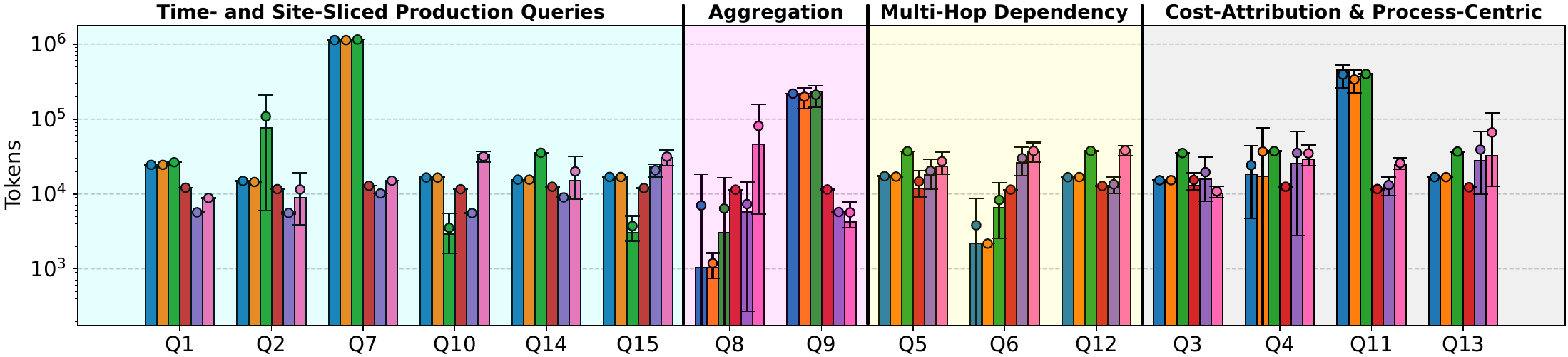}
  \vspace{-1.5em}
  \caption{Token cost: Bars indicate median, marker the mean and error bars the standard deviation from the mean.}
  \label{fig:eval-cost}
  \end{subfigure}
\vspace{-1.5em}
\caption{Results for the graph analytics queries. Details of individual queries and the explanation of query groups are in Appendix~\ref{sec:app:queries:industry}.}
\label{fig:eval-graph-analytics}
\vspaceSQ{-1em}
\end{figure*}

We now show advantages of \schemenameAS over the state of the art in three relevant metrics: success rates of solved graph tasks, running times, and token costs. These three metrics capture the three orthogonal dimensions that determine whether an LLM-enhanced graph system is practical at scale: (1) {task effectiveness}, (2) {systems efficiency}, and (3) {economic viability}.

\textbf{\uline{\schemenameAS Variants.} } 
We test three variants of \schemenameA: a \textbf{Basic \schemenameA} (includes the Semantic Catalog and the Adaptive Toolset), a \textbf{No Descriptors} variant (no Semantic Catalog), and a \textbf{Special Tools} variant (with task-specific operators, manually designed for selected complex queries). The last variant showcases how \schemenameAS supports extensibility through custom tool additions without altering the underlying framework.

\textbf{\uline{Comparison Baselines.} } 
We consider \textbf{LangChain}~\cite{Langchain}, the most widely adopted open-source framework for LLM-enhanced data access and analytics. LangChain serves as a strong baseline because it offers flexible prompt engineering, tool calling, and integration with graph query engines such as Neo4j~\cite{neo4j_book}. We consider three LangChain-based variants: \textbf{Basic LangChain} (it uses standard prompt templates that expose declarative schema structures but without any example instances), \textbf{Enhanced LangChain} that represents a best-effort optimization of LangChain prompts to maximize schema grounding and reduce ambiguity (it augments prompts with richly annotated schema information and includes synthetic example values), and \textbf{Enhanced LangChain + Semantic Catalog}.

\textbf{\uline{Datasets.} } 
\schemenameAS supports any labeled property graphs~\cite{besta2023graph, besta2023demystifying} or their variants such as heterogeneous graphs or knowledge graphs~\cite{besta2024hardware}. To showcase the usefulness of our design in real-world business settings, we harness an industry in-house graph with 230k vertices and 314k edges; we also use the \textsc{WikiDataSets--Countries} dataset~\cite{boschin2019wikidaatsets} with 3.3k vertices and 10.2k edges.

\textbf{\uline{Strict Token Budgets \& Small LLMs.} }
We use GPT-4o as the backend LLM (as opposed to heavier LLMs such as GPT-5) to keep reasoning accurate and affordable under strict token budgets; it also makes the system portable to other cost-efficient models (including small open models~\cite{hanke2024open, chen2024role, zhang2024tinyllama, bansal2024smaller}) without redesign.

\vspace{0.5em}
\noindent
Additional evaluation setup details are in Appendix~\ref{sec:app:setup}.

\subsection{Advantages for Industry Workloads}

\textbf{\uline{Success rate.} }
Figure~\ref{fig:eval-success} shows that \textbf{\schemenameS + Special Tools} successfully answers all queries correctly, substantially outperforming LangChain variants and the structure-only ablation; \textbf{Basic \schemename} is generally second-best, with clear advantages over all LangChain baselines on harder queries. In contrast, Basic and Enhanced LangChain achieve low to moderate success and degrade markedly on multi-hop, schema-heavy tasks (queries 10--15). Even when equipped with iterative prompting, LangChain frequently fails to {revise} its approach: across retries it often emits essentially the same query repeatedly rather than reflecting on execution feedback and generating meaningfully different candidates, which directly contributes to its poor performance on harder queries.

\textbf{\uline{Token usage.} }
In token usage (Figure~\ref{fig:eval-cost}) \textbf{\schemenameS + Special Tools} achieves \emph{consistently low and stable} token costs across query families, and is often \emph{comparable to or lower than} the LangChain baselines. In contrast, LangChain variants exhibit pronounced \emph{variance} with occasional large spikes on specific queries, reflecting longer in-context traces and repeated attempts to recover from brittle generations. \textbf{Basic \schemename} remains competitive but shows slightly higher dispersion than the Special Tools variant, while \textbf{\schemenameS - No Descriptors} typically falls in between. Overall, \schemenameAS improves \emph{cost-per-correct-answer} on both axes: it reduces failures while keeping per-run token costs competitive, yielding cheaper correct outputs on realistic workloads.

\textbf{\uline{Running time.} }
In end-to-end latency, the best-performing configuration is \textbf{\schemenameS + Special Tools}, which consistently yields the fastest response times. After pruning redundant tools from the toolkit, the system becomes noticeably more responsive: median per-query latency drops to the low single-digit seconds (typically $\approx$2--3\,s in our setup), and the agent remains interactive even when issuing multiple tool calls per query. Importantly, these additional steps do not translate into prohibitive delays. For instance, on queries Q3 and Q5 we observed cases where the agent initially selected a suboptimal tool or provided imperfect arguments, but then corrected itself in subsequent steps and still produced the correct answer after a short sequence of 2--3 tool invocations. This indicates that the proposed design supports self-correction as a first-class behavior while maintaining interactive response times.

By contrast, the LangChain-based baseline exhibits dramatically longer runtimes than even \textbf{Basic \schemename}, consistent with orchestration overhead and the cost of iterative retries, typically needing 2 minutes per query on average, corresponding to a $\sim$40--60$\times$ slowdown, which makes LangChain unsuitable for interactive use.

Overall, these results suggest a clear design guideline: a small, carefully optimized set of general-purpose tools with compact descriptions can achieve both lower latency and higher reliability than a large, highly specialized toolkit. Beyond reducing invocation overhead, tool pruning also narrows the agent's action space, which appears to improve tool selection and parameterization, thereby reducing wasted calls and further stabilizing end-to-end runtimes.

\subsection{Detailed Per-Query Analysis and Insights}

\textbf{\uline{When less context wins (Q1, Q8, Q9).} } 
\iftr
A notable pattern appears on queries \#1, \#8, and \#9.
\fi
Excessive semantic detail occasionally \emph{over-constrains} the agent: the semantic-heavy variants sometimes overuse available tools or fixate on narrower interpretations, increasing tokens without improving accuracy. For instance, on Q9 the \textbf{\schemenameS - No Descriptors} ablation study succeeds by leveraging the last-resort dynamic-Cypher tool directly. On the contrary, the semantic-rich variants persist with higher-level tools and incur overhead. A similar effect appears on Q8, where the semantic variant tries hard to route through specific tools. This suggests that \emph{more semantic context is not always better}; it must be gated by query type and uncertainty.

The success on Q9 is enabled by a feature of \schemenameAS in which it can (as a last resort, activated as an option after several failed attempts) completely bypass the compilation of the query from a pre-selected operator, and generate the Cypher query directly. 

\textbf{\uline{Semantics become essential (Q10–Q15).} } 
For queries 10--15, only models with semantic guidance solve the tasks reliably. These queries require disambiguating business entities, temporal qualifiers, and role-specific edges (e.g., linking processes to organizations in a specific year). Here, \emph{Semantic Descriptors significantly improve reasoning}, lifting success to near ceiling for \textbf{\schemenameS + Special Tools} and a lot for \textbf{Basic \schemename}, while LangChain variants plateau at low success.

\ifMORE
\subsection{Advantages Over Graph Learning Methods}

\maciej{Compare to GNNs / GFMs}
\fi

\subsection{Ablation Takeaways}

\textbf{\uline{Impact of Semantic Catalog.} }
Removing descriptors significantly reduces overall success, especially on disambiguation-heavy queries (Q10–Q15), but can occasionally {help} on simple list/lookup tasks by encouraging the agent to use the dynamic-Cypher tool directly (Q1, Q9). This motivates a fallback path where descriptors are switched off selectively if the task becomes sensitive to over-specification.

\textbf{\uline{Value of special-purpose tools.} } 
The \textbf{\schemenameS + Special Tools} variant demonstrates that \schemenameA's architecture supports \emph{domain-tuned accelerators} that both raise success and stabilize token usage. By collapsing multi-step plans into a single tool with targeted parameters, the agent avoids exploratory chains and redundant summaries, yielding consistent performance on hard queries.

\subsection{Advantages in the WikiDataSets Benchmark}

We evaluate the \textsc{Countries} benchmark of \textsc{WikiDataSets}~\cite{boschin2019wikidaatsets}, a standardized, topic-specific subgraph of Wikidata that preserves real-world schema heterogeneity (multiple administrative-unit types, aliases, and nested \emph{part-of}/\emph{located-in} relations) while remaining small enough for controlled, reproducible NL-to-graph analytics experiments (detailed in Appendix~\ref{sec:app:setup}).

The results are shown in Figure~\ref{fig:eval-countries} (queries Q1--Q15 are detailed in Appendix~\ref{sec:app:queries:countries}). \schemenameAS attains near-ceiling success across essentially all queries, while LangChain-style baselines fail on many prompts and exhibit high variance across question types. The gains are most pronounced on queries that require mapping NL terms to the correct administrative type (e.g., \emph{province} vs.\ \emph{state} vs.\ \emph{county} vs.\ \emph{federal district} vs.\ \emph{constituent country}) and then executing part-of/located-in--whole reasoning reliably; these are exactly the failure modes of direct NL-to-query generation.

\textbf{\uline{Failure mode analysis.} }
We observe that prompt-only baselines often pick the wrong \emph{semantic target} (treating the entity as a country rather than an administrative unit, or confusing the expected return type e.g., answering a \emph{type} question with a \emph{location} and vice versa. Second, they conflate near-synonyms and Wikidata-specific unit categories (e.g., mapping ``federal district'' to a ``state''-like category, or mixing ``province'' and ``region''). Third, they mishandle directionality in containment predicates (swapping \emph{part-of} vs.\ \emph{has part}, yielding inverted traversals). Finally, multi-hop chains prematurely terminate after a plausible intermediate node (returning the immediate parent rather than the requested enclosing political unit, especially when multiple containment paths exist).

\textbf{\uline{Token usage.} }
When using \textbf{Basic \schemename}, the token usage remains stable and comparatively low whenever the agent reliably grounds to the database runtime: the \emph{overall} median token count is $\approx$29.7k tokens per query, with most successful runs clustering tightly around $\approx$29--31k tokens when the agent performs 2--4 database-backed tool calls. The main deviations are informative rather than noise: (i) unusually \emph{low} token counts (e.g., $\approx$7k or $\approx$14k) typically correspond to cases where the LLM stopped consulting the database and instead relied on ``insight''/prior knowledge after a failure or partial execution, which is precisely when correctness can degrade; and (ii) unusually \emph{high} token counts (peaking at $\approx$55k) occur on harder queries that trigger longer corrective sequences, including occasional 6-call trajectories. Consistent with this interpretation, the descriptor-grounded \textbf{Basic \schemename} variant achieves a high aggregate success rate (with the few sub-ceiling cases concentrated in the most schema-ambiguous prompts, e.g., Q3 at 0.8 and Q4/Q8/Q9 at 0.9), while keeping the majority of successful runs in the same narrow token band.

\textbf{\uline{Running time.} }
\textbf{Basic \schemename} stays in the same interactive regime as discussed earlier: end-to-end latency remains on the order of a few seconds per query (typically $\approx$3--4\,s in our setup), even when it executes multiple tool calls.

Taken together, these results reinforce the core design point: Semantic Descriptors help select the correct schema element/operator, while the Non-LLM runtime performs the expensive traversal and filtering deterministically; this combination yields both high accuracy and predictable, interactive cost/latency profiles compared to prompt-only baselines.
\schemenameAS also uses substantially fewer tokens per query: by delegating database-grade execution to a Non-LLM runtime and returning only compact summaries, it avoids long in-context intermediate traces and repeated re-querying observed in baselines, yielding lower token cost while improving accuracy.

\begin{figure}[t]
  \centering
  \includegraphics[width=1.0\columnwidth]{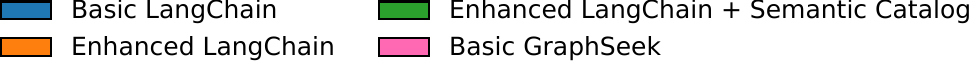}
  \begin{subfigure}[t]{\columnwidth}
  \includegraphics[width=1.0\columnwidth]{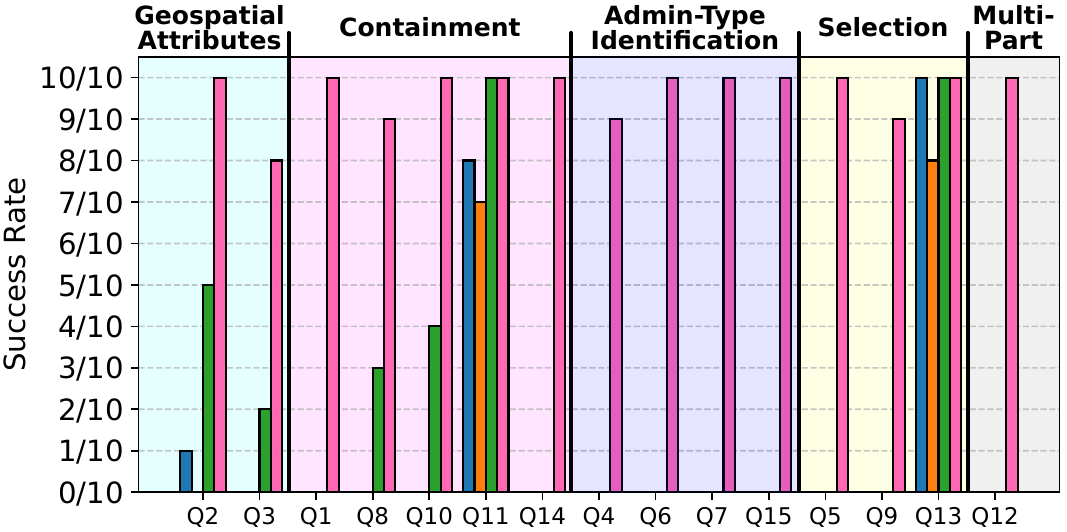}
  \vspace{-1.5em}
  \caption{Success rate.}
  \label{fig:eval-countries-success}
  \end{subfigure}
  \begin{subfigure}[t]{\columnwidth}
  \includegraphics[width=1.0\columnwidth]{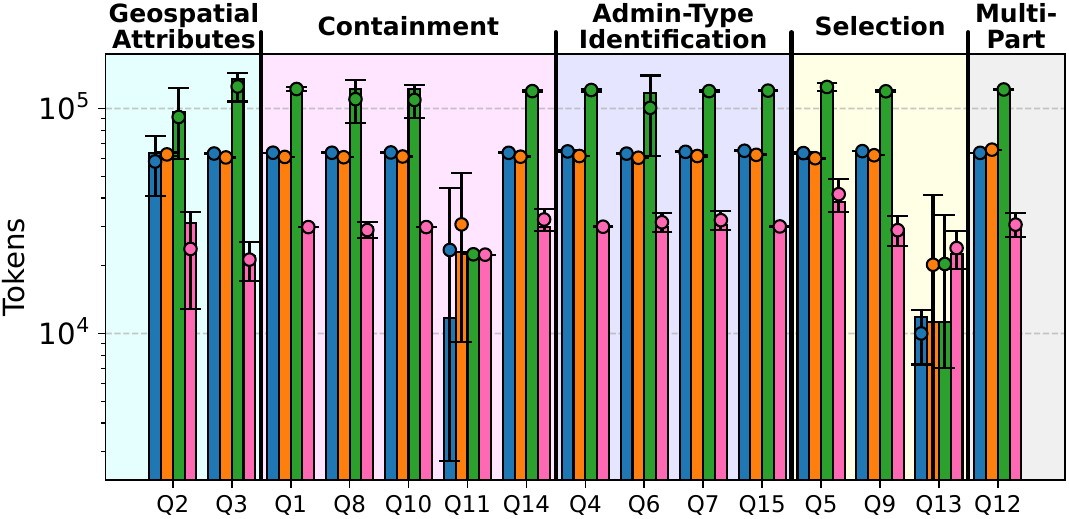}
  \vspace{-1.5em}
  \caption{Token cost: Bars indicate median, marker the mean and error bars the standard deviation from the mean.}
  \label{fig:eval-countries-cost}
  \end{subfigure}
\vspace{-1.0em}
\caption{Results for the \textsc{Countries} benchmark of \textsc{WikiDataSets}. Details of queries and their groups are in Appendix~\ref{sec:app:queries:countries}.}
\label{fig:eval-countries}
\ifconf
\else
\vspaceSQ{-1.5em}
\fi
\end{figure}

\iftr\section{RELATED WORK}

\iftr
Our work touches upon numerous domains. In Section~\ref{sec:landscape}, we have already described the relationship between \schemenameAS and \textbf{graph representation learning}, general \textbf{LLM-enhanced data analytics} beyond graphs, and recent works sitting at the overlap of \textbf{prompt engineering and graphs}. We now discuss other related areas.
\else
Our work touches upon numerous domains. In Appendix~\ref{sec:landscape} and in Figure~\ref{fig:landscape}, we describe the relationship between \schemenameAS and \textbf{graph representation learning}, general \textbf{LLM-enhanced data analytics} beyond graphs, and recent works sitting at the overlap of \textbf{prompt engineering and graphs}. We now discuss other related areas.
\fi

\textbf{Graph databases. }
A mature ecosystem of graph databases supports property graphs and RDF/OWL, optimized for fast adjacency accesses and transactional workloads, which are capabilities that LLM-enhanced analytics can {invoke} rather than re-implement~\cite{angles2018introduction, besta2023demystifying, Angles2012, davoudian2018survey, han2011survey, gajendran2012survey, gdb_survey_paper_Kaliyar, kumar2015domain}. Representative systems include Neo4j~\cite{neo4j_book}, Graph Database Interface~\cite{besta2023gdi, besta2023demystifying}, TigerGraph~\cite{tigergraph2022ldbc, tiger_graph_links}, Amazon Neptune~\cite{amazon_neptune_links}, JanusGraph~\cite{janusgraph}, Virtuoso~\cite{virtuoso_links}, ArangoDB~\cite{arangodb_links}, RedisGraph~\cite{redisgraph_links}, Dgraph~\cite{dgraph_links}, Stardog~\cite{stardog_links}, and GraphDB~\cite{graphdb_links}, among many others~\cite{cge_paper, chen2019grasper, azure_cosmosdb_links, apache_jena_tbd_links, oracle_spatial, tesoriero2013getting, profium_sense_links, triplebit_links, gbase_paper, graphbase_links, besta2025higher, graphflow, livegraph, memgraph_links, dubey2016weaver, sparksee_paper, allegro_graph_links, mormotta_links, brightstardb_links, gstore, anzo_graph_links, datastax_links, blaze_graph_links, besta2023graph, besta2022neural, cayley_links}. In LLM-enhanced graph analytics, these engines serve as the \emph{execution substrate}: the LLM plans, grounds user intent to schema, and compiles to Cypher/Gremlin/SPARQL-like queries; the DBMS then executes at scale with ACID/ACL semantics. Our approach follows this separation of concerns by delegating physical execution to a non-LLM executor while keeping prompts small and schema-aware.

\textbf{Graph analytics frameworks and schemes. }
\iftr
High-perfor-mance
\else
High-performance
\fi
graph mining and analytics stacks provide scalable primitives (e.g., traversals, motif finding, community detection) and domain-specific methods that LLM agents can orchestrate. Examples include GraphMineSuite (GMS)~\cite{gms}, ProbGraph~\cite{besta2022probgraph}, and a large body of systems/algorithms spanning parallel traversal, frequent subgraph mining, dynamic/streaming updates, coloring, communication-avoiding kernels, and numerous others~\cite{satish2014navigating, yan2017big, sakr2021future, besta2015accelerating, besta2017slimsell, besta2017push, besta2019slim, besta2020highcolor, besta2020substream, besta2021sisa, besta2019practice, chakrabarti2006graph, choudhury2017nous, cook2006mining, jiang2013survey, horvath2004cyclic, duan2019dynamic, gianinazzi2018communication, gianinazzi2021parallel, gianinazzi2021learning, strausz2022asynchronous, bernstein2021deamortization, neiman2015simple, solomonik2017scaling, qin2019mining}. Concretely, such systems expose composable operators; our LLM layer can call such operators as tools, letting the backend handle execution details.

\textbf{Agent systems. }
The landscape of {agent systems} has emerged during the last years, with example designs such as Reflexion~\cite{shinn2024reflexion}, AutoGen~\cite{wu2023autogen}, MetaGPT~\cite{hong2023metagpt}, Knowledge Graph of Thoughts~\cite{besta2025affordable}, and many others~\cite{Langchain, rush2023minichain, kim2023llm, sumers2023cognitive, gaia, guo2024large, edge2024local, besta2024demystifying, zhuge2024language, beurer2023prompt, kagaya2024rap, zhao2024expel, stengel2024regal, zhu2024knowagent, chen2023autoagents, liu2023dynamic, besta2024checkembed, besta2024graph, chu2024professional}. These schemes have been applied to numerous tasks in reasoning~\cite{creswell2022selection, bhattacharjya2024foundation, besta2024demystifying}, planning~\cite{wang2023describe, prasad2023adapt, shen2024hugginggpt, huang2024anpl}, software development~\cite{tang2024collaborative}, and others~\cite{xie2023openagents, schick2023toolformer, beurerlarge}.
The difference to our work is that \schemenameAS can be seen as the first agent system that offers LLM-enhanced interactive graph analytics.

\textbf{Retrieval-Augmented Generation (RAG). }
RAG connects LLM prompts to external stores (documents, tables, vector indexes, or graph repositories) to reduce hallucinations and compress context. Designs range from classic document RAG to adaptive and hierarchical retrieval~\cite{gao2024retrieval, zhao2024retrieval, hu2024rag, huang2024survey, yu2024evaluation, mialon2023augmented, li2022survey, abdallah2024generator, delile2024graph, manathunga2023retrieval, zeng2024federated, wewer2021updating, xu2024active, sarthi2024raptor, asai2023selfrag, yu2023chain, gutierrez2024hipporag, edge2024local, besta2024multi, llamaindex}. For graph analytics, we use RAG to retrieve schema snippets and tool documentation, fetch compact graph summaries (neighborhood sketches, degree/label histograms), and incorporate prior conversation artifacts; all are injected as fixed-size features into prompts, preserving small-context compatibility.

\textbf{Task decomposition \& planning. }
Decomposition turns complex graph questions into executable subgoals (e.g., “identify candidates,” “filter by constraints,” “aggregate KPIs”). Methods such as ADaPT~\cite{prasad2023adapt}, ANPL~\cite{huang2024anpl}, HuggingGPT~\cite{shen2024hugginggpt}, and graph-based control-flow encodings (ScrapeGraphAI~\cite{scrapegraph-ai}, LangGraph~\cite{langgraph}) provide templates for converting natural instructions into DAGs of tool invocations. Our system adopts iterative planning with schema/tool grounding, allowing safe backtracking: failed subgoals trigger verified re-plans rather than longer prompts.

\textbf{Building and reusing past experience. }
Persistent memories and experience bases capture successful tool chains, schemas, and failure modes. RAP~\cite{kagaya2024rap}, ExpeL~\cite{zhao2024expel}, ReGAL~\cite{stengel2024regal}, Hypotheses-to-Theories~\cite{zhu2024large}, KnowAgent~\cite{zhu2024knowagent}, and Reflexion~\cite{shinn2024reflexion} demonstrate how episodic/task memories, action knowledge bases, and reflective critiques improve stability and sample efficiency. In our system, we use the past history of messages within a given task to tune the next data reasoning step.

\textbf{Fine-tuning for graph analytics. }
Graph-specific instruction tuning and reinforcement learning (RL) can specialize LLMs for tool use and structural reasoning. GraphTool-Instruction~\cite{wang2025graphtool} fine-tunes for decomposition and tool invocation on graph tasks, while GraphChain~\cite{wei2025graphchain} applies reinforcement learning to learn tool-chaining policies at scale. More broadly, these works exemplify \emph{parametric} specialization for graph reasoning that needs training. In contrast, our scheme targets \emph{inference-time} adaptivity with no fine-tuning, emphasizing plug-and-play deployment, token efficiency, and portability across small-context models.

\textbf{Multi-agent collaboration. }
Coordination among multiple experts (planners, verifiers, explainers) can yield robustness and throughput. CodeAgent~\cite{tang2024collaborative}, AutoAgents~\cite{chen2023autoagents}, DyLAN~\cite{liu2023dynamic}, Professional Agents~\cite{chu2024professional}, MBTI-in-Thoughts~\cite{besta2025mbti}, and AutoGen~\cite{wu2023autogen} study division of labor and communication protocols; Agent Forest~\cite{li2024more} uses sampling/voting, while GPTSwarm~\cite{zhuge2024language} models agents as computational graphs. Although our current system achieves high fidelity with a single orchestrator, these strategies are complementary for future scaling (e.g., parallel exploration of alternative graph plans with cross-checking).

\textbf{Tool use and adaptivity. }
Tool-augmented LLMs extend beyond text-only reasoning to call external programs and databases~\cite{schick2023toolformer, xie2023openagents, beurerlarge}. ToolNet~\cite{liu2024toolnet} models multi-tool application as a directed graph to schedule iterative tool use at scale. Critically for evolving enterprise graphs, \emph{non-static} tools that can be revised at inference time (e.g., DynaSaur~\cite{nguyen2024dynasaur}) enable the agent to synthesize/modify operators as schemas drift. {Our system embraces this paradigm: we maintain a self-adaptive toolset that the LLM can refine (parameterize, compile, or swap) on demand, while confining prompts to constant-size schema/tool descriptors and delegating heavy computation to the executor.}

\fi
\section{CONCLUSION}

We present \schemenameA, a framework for large-scale LLM-enhanced graph analytics that employs a principled separation between a \emph{Semantic Plane} and an \emph{Execution Plane}. In the Semantic Plane, the LLM agent plans and interprets analytics requests by reasoning over compact, execution-derived summaries and a \emph{Semantic Catalog} that annotates both schema and available operators; in the Execution Plane, a deterministic Non-LLM Executor performs data access, filtering, and post-processing. Complementing this core design, \schemenameAS supports an \emph{Adaptive Toolset} whose operator implementations can evolve at inference time, ensuring robustness against evolving data at controlled (and often reduced) token budgets.

Looking forward, the Semantic Catalog + plane separation generalizes beyond property graphs: to tabular and relational databases, time-series pipelines, heterogeneous modalities (documents, logs, events), and streaming settings such as dynamic graphs where summaries must be incremental. Deeper integration with database engines (e.g., cost-aware compilation), principled policies for catalog construction, and learned budget controllers for operator selection are promising directions. We believe \schemenameAS provides an operational foundation for scalable, trustworthy LLM-based analytics over evolving, industry-scale irregular datasets.



\maciej{LATENCY / RUNNING TIMES!!!} \orest{DONE}

\maciej{Double check with comments here \url{https://openreview.net/forum?id=WqNDuWyoOL\#discussion}}

\maciej{' My biggest concern is about the motivation. It is unclear to me what real applications on graphs allow hallucination risks}

\maciej{'Finally, the absence of released code or data limits reproducibility.'}

\maciej{Advantages Across Different LLMs}

\maciej{Try different temperatures?}

\maciej{'It is also unclear what statistics the framework extracts from the graphs (in particular from the experiments)'}

\ifnonbd
\begin{acks}
We thank Hussein Harake, Colin McMurtrie, Mark Klein, Angelo Mangili, and the whole CSCS team granting access to the Ault and Alps machines, and for their excellent technical support.
We thank Timo Schneider for help with infrastructure at SPCL.
This project received funding from the European Research Council (Project PSAP, No.~101002047), and the European High-Performance Computing Joint Undertaking (JU) under grant agreement No.~955513 (MAELSTROM). This project was supported by the ETH Future Computing Laboratory (EFCL), financed by a donation from Huawei Technologies. This project received funding from the European Union's HE research and innovation programme under the grant agreement No. 101070141 (Project GLACIATION).
We gratefully acknowledge Polish high-performance computing infrastructure PLGrid (HPC Center: ACK Cyfronet AGH) for providing compute facilities within computational grant no.~PLG/2024/017103. We acknowledge the Swiss AI Initiative for the computational grant.
\end{acks}
\fi

\bibliographystyle{ACM-Reference-Format}
\bibliography{references.complete}

\appendix

\clearpage
\section{SYSTEM DESIGN: ADDITIONAL DETAILS}
\label{sec:app:arch}

We provide additional details on the system design.

\subsection{Example Semantic Descriptors}

\begin{figure}[!h]
\vspace{-1.0em}
 \centering
 \includegraphics[width=0.48\textwidth]{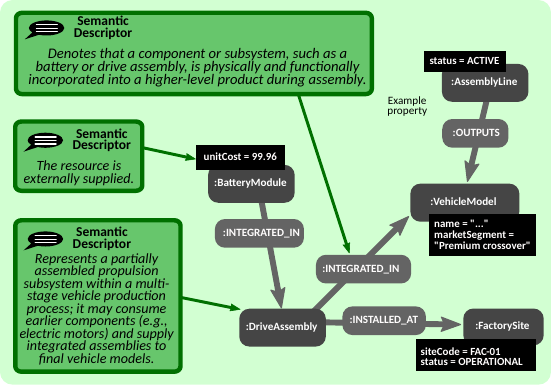}
\vspace{-2.0em}
\caption{Example anonymized schema, together with example Semantic Descriptors, of the used graph dataset.}
\label{fig:graph}
\vspace{-2.0em}
\end{figure}

We showcase example Semantic Descriptors and their associated schema elements in Figure~\ref{fig:graph}.

\begin{figure*}[t]
 \centering
 \includegraphics[width=1.0\textwidth]{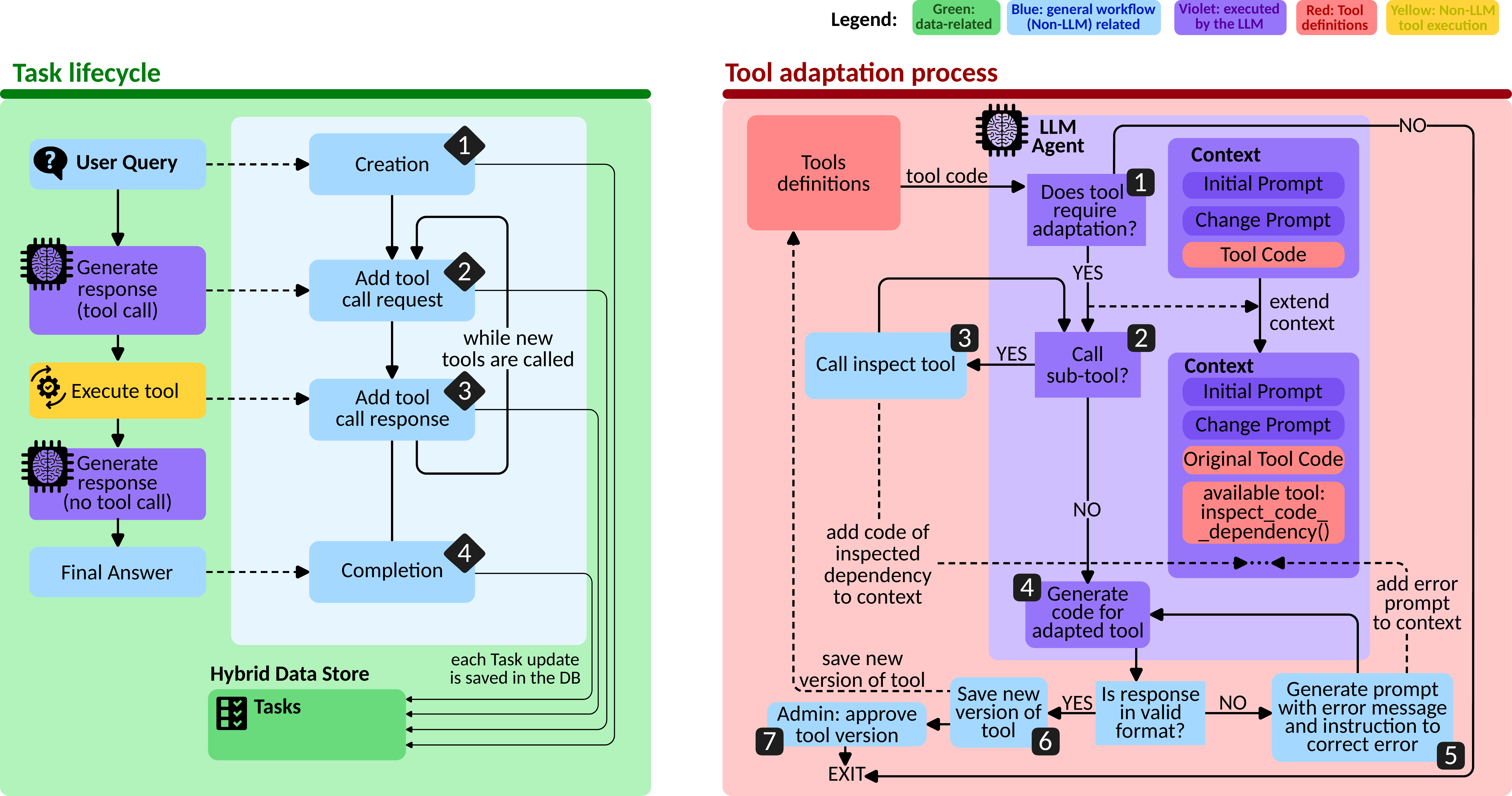}
 \vspace{-2em}
\caption{Overview of the workflows for the task lifecycle and the tool adaptation.}
\label{fig:overview-additional}
\vspace{1em}
\end{figure*}

\subsection{Task Lifecycle}
\label{sec:task-lifecycle}

The task lifecycle is pictured on the left side of Figure~\ref{fig:overview-additional}.
Each user interaction in \schemenameAS is encapsulated within a \textsf{Task}, which serves as a structured container for execution state, tool usage, and result traceability.
%
%
Upon receiving a new user request $x$, the system instantiates a new Task object~\includegraphics[scale=0.15,trim=0 16 0 0]{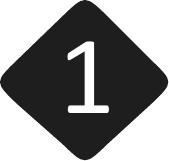}. As the LLM Agent processes the request, every tool invocation is logged within the Task~\includegraphics[scale=0.15,trim=0 16 0 0]{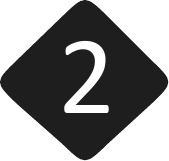}, along with a reference to its corresponding result stored in the Hybrid Data Store~\includegraphics[scale=0.15,trim=0 16 0 0]{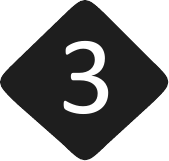}.

This design allows the Task to act as an execution trace: it records not only the sequence of tool calls but also links to intermediate and final outputs, including graphs, tables, or summaries. Once the LLM produces the final response, the Task is marked as completed and becomes immutable, ready for future auditions~\includegraphics[scale=0.15,trim=0 16 0 0]{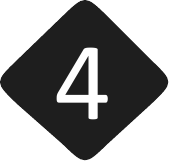}.

\subsection{Tool Adaptation Process}
\label{sec:tool-adaptation}

The tool adaptation is pictured on the right side of Figure~\ref{fig:overview-additional}.
\schemenameAS supports dynamic tool adaptation via an inference-time, LLM-driven code synthesis loop.
%
%
When a schema changes, the LLM Agent may inspect a tool (including its utility call chain)~\includegraphics[scale=0.2,trim=0 16 0 0]{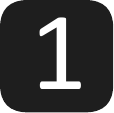}--\includegraphics[scale=0.2,trim=0 16 0 0]{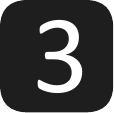}, propose a revised template or implementation consistent with the operator signature declared in the Semantic Catalog, and submit a candidate. The LLM Agent then generates a new version of the tool~\includegraphics[scale=0.2,trim=0 16 0 0]{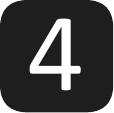}. This candidate is automatically validated against a strict schema (defined with JSON Schema or any other specification form) provided in the system prompt. If validation fails, the LLM Agent is tasked with re-generating the code in light of the returned error message~\includegraphics[scale=0.2,trim=0 16 0 0]{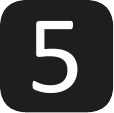}. Validated code is stored as an \emph{unverified} version~\includegraphics[scale=0.2,trim=0 16 0 0]{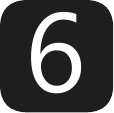} and, after lightweight human approval, promoted to the live library~\includegraphics[scale=0.2,trim=0 16 0 0]{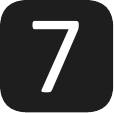}.

\subsection{Example Compact Summary}

An example compact $O(1)$ graph summary is shown in Listing~\ref{lst:graph-summary}. It depicts constant-size metadata for a subgraph centered on a single EV configuration. The fields \textsf{summary}, \textsf{preview\_graph\_stats} (aggregate statistics), and a fixed-size \textsf{preview\_graph} together describe a local neighborhood around a vehicle model (e.g., \textsf{Model Z3}), capturing semantically typed nodes such as \textsf{BatteryModule}, \textsf{DriveAssembly}, \textsf{VehicleModel}, \textsf{FactorySite}, and \textsf{AssemblyLine}, along with key relationships including \textsf{INTEGRATED\_IN}, \textsf{OUTPUTS}, \textsf{INSTALLED\_AT}, and \textsf{CONNECTED\_TO}. The full subgraph remains addressable via a \textsf{graph\_handle}, enabling the LLM to reason over this compact preview while offloading deeper traversal or analytics to the Non-LLM Executor.

\begin{figure*}[t]
\begin{lstlisting}[caption={An example compact graph summary and its previous use in the LLM context within \schemenameA.}, showstringspaces=false, belowskip=1em, label={lst:graph-summary}, basicstyle=\tt\small]
{
  "summary": "Assembly plan for VehicleModel 'Model Z3' (key 'EV-Z3') at integration tier 'Tier-0' under cost class 'Actuals' for calendar year 2023 (Jan-Dec).",
  "graph_handle": "graph_ev_z3",
  "preview_graph_stats": {
    "nodes": 37,
    "relationships": 33,
    "node_types": ["BatteryModule", "DriveAssembly", "VehicleModel", "FactorySite", "AssemblyLine"],
    "relationship_types": ["INTEGRATED_IN", "OUTPUTS", "INSTALLED_AT", "CONNECTED_TO"]
  },
  "preview_graph": {
    "nodes": [
      { "id": "n1", "type": "VehicleModel",   "name": "Model Z3",  "key": "EV-Z3" },
      { "id": "n2", "type": "AssemblyLine",   "name": "Line A17" },
      { "id": "n3", "type": "BatteryModule",  "name": "Pack B90" },
      { "id": "n4", "type": "DriveAssembly",  "name": "Motor D21" },
      { "id": "n5", "type": "FactorySite",    "name": "Site Alpha" }
    ],
    "relationships": [
      { "id": "e1", "type": "INSTALLED_AT",  "source": "n2", "target": "n5" },
      { "id": "e2", "type": "OUTPUTS",       "source": "n2", "target": "n1" },
      { "id": "e3", "type": "INTEGRATED_IN", "source": "n3", "target": "n2" },
      { "id": "e4", "type": "CONNECTED_TO",  "source": "n4", "target": "n1" }
    ]
  },
  "notes": [
    "This is a constant-size (O(1)) graph summary and fixed-size preview.",
    "Use 'graph_handle' to retrieve the full subgraph or specific nodes/edges when required."
  ]
}
\end{lstlisting}
\end{figure*}

\section{EVALUATION SETUP: FURTHER DETAILS}
\label{sec:app:setup}

\subsection{Models}

All experiments use OpenAI's \texttt{GPT-4o} model. We enforce strict prompt budgets: a \emph{per-message} cap of 20{,}000 tokens for any single user input (never reached in our workloads) and a \emph{conversation history} budget of 10{,}000 tokens, after which older content is summarized or pruned as described in Section~\ref{sec:arch}.


\subsection{Datasets \& Benchmarks}

The underlying processing plant property graph slice contains 229{,}551 nodes, 313{,}635 relationships, and 20 distinct property keys, representative of an industry-scale workload for our queries and ablation studies.

The \textsc{WikiDataSets--Countries} dataset~\cite{boschin2019wikidaatsets} and benchmark is a topic-specific knowledge-graph subgraph extracted from Wikidata, where entities are included if they are instances of \emph{country} (Wikidata topic entity) or any subclass thereof (e.g., historical countries), yielding a labeled multirelational graph with facts of the form $(h,r,t)$. In addition to edges between selected nodes, the dataset retains facts from selected nodes to non-selected Wikidata entities as \emph{node attributes}, enabling queries that mix inter-country relations (e.g., borders, diplomatic relations) with property-style information (e.g., currency, capital). The released artifact includes separate files for edges, attributes, and dictionaries mapping entity/relation IDs to human-readable labels, supporting reproducible loading and evaluation across systems. The published metadata for \textsc{Countries} reports $3{,}324$ nodes and $10{,}198$ edges, with $48$ distinct relation types, plus $17{,}965$ distinct attributes across $37{,}096$ attribute facts (and $145$ distinct attribute-relation types).

\ifMORE
Second, we also consider the Open Academic Graph Version 3.1~\cite{tang2008arnetminer, zhang2019oag, zhang2023oag}, which incorporates the established MS Knowledge Graph (MAG)~\cite{wang2020microsoft}, with over 153M vertices and 422M edges. 
\fi

%

\subsection{Metrics \& Success Rate}

Each repeat is a fresh run in which the LLM Agent may internally issue any number of tool calls or self-corrections; the default value we use is up to 10. For LangChain, similarly, each repeat consists of up to 10 attempts within a single conversation, where each attempt sees the full history. A repeat is considered successful if any attempt yields a correct final answer, i.e., the LLM asserts that its graph query is correct, the query executes successfully, and the resulting output allows the model to produce a response consistent with the ground truth. Numerical tasks require exact correctness; tasks where certain tuples must be fetched are deemed executed correctly if at least one of the needed tuples has been retrieved.



\subsection{Workloads}

We evaluate \schemenameAS on a suite of 30 queries that exercise this model end-to-end, spanning period-based listings, multi-hop impact analysis (e.g., shortages), schema-aware entity/process linkages, and cross-modal presentation (tables/graphs). All queries are listed in Appendix~\ref{sec:app:queries}. For each query and method, we run \textbf{10 independent repeats} with the same LLM backend and database.
\section{USED GRAPH QUERIES: DETAILS}
\label{sec:app:queries}

\subsection{Industry Dataset}
\label{sec:app:queries:industry}

We evaluate on the following high-level graph queries (reformulated from the original domain to avoid disclosing collaborator-specific schema details):

\begin{description}[itemsep=0.2em]
    \item[Q1] List all battery modules produced in a given calendar year.
    \item[Q2] List all vehicle models assembled during a specified quarter of a given year.
    \item[Q3] Identify vehicle models built using a specified assembly-process family (given by a human-readable label).
    \item[Q4] Report aggregated costs of auxiliary modules (e.g., scrap or side assemblies) at a specified assembly tier during a given year.
    \item[Q5] Given a particular component module, determine which vehicle models would be impacted by a shortage.
    \item[Q6] If a given component module becomes unavailable, determine which factory sites and regions would be affected.
    \item[Q7] List all battery modules produced at a specified factory site.
    \item[Q8] Find vehicle models that are currently manufactured at three or more factory sites.
    \item[Q9] Find vehicle models scheduled for production in two or more distinct time periods.
    \item[Q10] For a given vehicle model identifier and month/year, list the factory sites scheduled to assemble it.
    \item[Q11] For a given vehicle model, identify component modules with the greatest contribution to total cost and/or output.
    \item[Q12] Determine which vehicle models are planned to include a specified component module.
    \item[Q13] Identify assembly processes associated with a specified business unit during a given year.
    \item[Q14] Enumerate all assembly processes executed at a specified factory site.
    \item[Q15] For a given vehicle model and quarter, list the assembly processes used to manufacture it.
\end{description}

For analysis, we group the queries into four families that exercise different graph-structured access patterns and stress distinct aspects of the underlying graph engine (e.g., neighborhood expansion, multiway joins, and aggregation).

\subsubsection{\textbf{Time- and Site-Sliced Production Queries.}}

These queries primarily perform attribute-filtered selection on nodes (e.g., vehicle models, battery modules, assembly processes), combined with short-range neighborhood traversal to calendar and factory-site nodes. They correspond to “slicing” the production graph along time and location dimensions.

\begin{description}[itemsep=0.2em]
    \item[Q1] List all battery modules produced in a given calendar year.
    \item[Q2] List all vehicle models assembled during a specified quarter of a given year.
    \item[Q7] List all battery modules produced at a specified factory site.
    \item[Q10] For a given vehicle model identifier and month/year, list the factory sites scheduled to assemble it.
    \item[Q14] Enumerate all assembly processes executed at a specified factory site.
    \item[Q15] For a given vehicle model and quarter, list the assembly processes used to manufacture it.
\end{description}

Structurally, these are 1–2 hop traversals over a bipartite or tripartite subgraph linking \emph{product} or \emph{process} nodes to \emph{calendar} and \emph{site} nodes. The queries start from one of these axes (time, site, or vehicle) and fan out over a small star-shaped neighborhood to recover the matching nodes. From the engine's perspective, this family stresses efficient indexed access on attributes (e.g., year, quarter), expansion over low-diameter patterns, and simple projections back to domain entities (modules, models, processes) that a downstream analytics pipeline would consume.

\subsubsection{\textbf{Multi-Site and Multi-Period Neighborhood Aggregation.}}

These queries introduce degree constraints and aggregation over distinct neighbors, moving beyond simple selection to counting structures in the graph.

\begin{description}[itemsep=0.2em]
    \item[Q8] Find vehicle models that are currently manufactured at three or more factory sites.
    \item[Q9] Find vehicle models scheduled for production in two or more distinct time periods.
\end{description}

\sloppy
In graph terms, they compute degree-based predicates over the vehicle–site and vehicle–time bipartite relations (e.g., $\deg_{\text{site}}(\text{vehicle}) \geq 3$ or $\deg_{\text{time}}(\text{vehicle}) \geq 2$). This requires grouping by vehicle nodes and aggregating over incident edges to distinct neighbors, which can be seen as a constrained neighborhood expansion followed by a \emph{group-by} on the center node. Such queries are representative of capacity planning and risk-spreading analyses in EV manufacturing (e.g., “which models are geographically diversified”), while also stressing the graph engine’s ability to perform combiners and degree computations efficiently at scale.

\subsubsection{\textbf{Dependency and Impact-Propagation Queries.}}

These queries follow multi-hop dependencies (e.g., component module $\rightarrow$ assembly $\rightarrow$ vehicle model $\rightarrow$ site/region).
  
\begin{description}[itemsep=0.2em]
    \item[Q5] Given a particular component module, determine which vehicle models would be impacted by a shortage.
    \item[Q6] If a given component module becomes unavailable, determine which factory sites and regions would be affected.
    \item[Q12] Determine which vehicle models are planned to include a specified component module.
\end{description}
  
They correspond to reachability and constrained neighborhood-expansion queries over two coupled subgraphs: the product-structure graph and the deployment graph (vehicles to sites/regions). Typical patterns here are 2–4 hop paths with branching (a component can appear in multiple assemblies, each used in multiple vehicles, deployed at multiple factories). This family captures “what-if” reasoning that may be difficult to express compactly in flat relational schemas but natural in a graph setting, and it stresses the ability of the system to perform multi-hop traversals with selective pruning and to propagate impact sets across large neighborhoods.

\subsubsection{\textbf{Cost Attribution and Process/Organization-Centric Views.}}

These queries combine graph traversal with aggregation, ranking, and alternative “root” perspectives on the same underlying topology.
  
\begin{description}[itemsep=0.2em]
    \item[Q3] Identify vehicle models built using a specified assembly-process family (given by a human-readable label).
    \item[Q4] Report aggregated costs of auxiliary modules (e.g., scrap or side assemblies) at a specified assembly tier during a given year.
    \item[Q11] For a given vehicle model, identify component modules with the greatest contribution to total cost and/or output.
    \item[Q13] Identify assembly processes associated with a specified business unit during a given year.
\end{description}
  

\subsection{\textsc{WikiDataSets--Countries} Benchmark}
\label{sec:app:queries:countries}

We evaluate on the following high-level graph queries:

\begin{description}[itemsep=0.2em]
    \item[Q1] Where is the state of Morelos located?
    \item[Q2] In which hemisphere is Botswana located?
    \item[Q3] Which country is located in the Northern Hemisphere?
    \item[Q4] What type of area is Mexico City in Mexico?
    \item[Q5] Which one is a state in Australia?
    \item[Q6] New South Wales is an Australian what?
    \item[Q7] What type of subdivision is Alabama in the United States of America?
    \item[Q8] Where is California located?
    \item[Q9] Which federal district is located in the United States?
    \item[Q10] In which country is Yukon located?
    \item[Q11] Cornwall is a county located in which country?
    \item[Q12] England is a constituent of which political unit, along with which other parts?
    \item[Q13] Which type of subdivision is found in England?
    \item[Q14] Heilongjiang is a province located in which country?
    \item[Q15] Which type of subdivision is Xinjiang in the People's Republic of China?
\end{description}

For analysis, we group the \textsc{Countries} queries into five families that exercise different graph-structured access patterns over an administrative-division KG (e.g., type constraints, part--whole joins, and attribute-based predicates such as hemisphere). Throughout, we view the dataset as a labeled multirelational graph with entity nodes and edges (facts) of the form $(h,r,t)$ plus attribute facts $(v,a,\alpha)$, where $\alpha$ is a literal (e.g., \texttt{Northern Hemisphere}).

\subsubsection{\textbf{Geospatial Attribute Predicates (Hemisphere).}}

These queries test attribute-level reasoning about geographic properties.

\begin{description}[itemsep=0.2em]
  \item[Q2] In which hemisphere is Botswana located?
  \item[Q3] Which country is located in the Northern Hemisphere?
\end{description}

Graph-theoretically, this family is dominated by \emph{attribute lookup} and \emph{attribute filtering}. Q2 is a single-node attribute read $(\texttt{Botswana},\texttt{hemisphere},\alpha)$, whereas Q3 is an inverse selection $\{v : (v,\texttt{hemisphere},\texttt{Northern Hemisphere})\}$ possibly followed by a type constraint $v \in \textsf{Country}$. From the engine's perspective, these queries stress efficient access to literal-valued properties and simple selection predicates, rather than deep traversal.

\subsubsection{\textbf{Containment / Located-In Joins.}}

These queries ask for the containing political unit of a named place.

\begin{description}[itemsep=0.2em]
  \item[Q1] Where is the state of Morelos located?
  \item[Q8] Where is California located?
  \item[Q10] In which country is Yukon located?
  \item[Q11] Cornwall is a county located in which country?
  \item[Q14] Heilongjiang is a province located in which country?
\end{description}

Structurally, these are \emph{short-path part--whole} queries, typically a 1-hop join along a containment predicate such as \texttt{locatedIn} / \texttt{partOf}. In abstract form, they compute $t$ such that $(v,\texttt{locatedIn},t)$ (or $(v,\texttt{partOf},t)$) for a given $v$. Depending on the KG schema, $t$ may be a country or an intermediate administrative unit, yielding a small number of additional hops (e.g., $v \rightarrow \textsf{State} \rightarrow \textsf{Country}$). This family stresses reliable relation selection (choosing the correct containment edge among many plausible predicates) and robust entity linking for administrative units.

\subsubsection{\textbf{Administrative-Type Identification.}}

These queries ask for the administrative \emph{type} of a named entity (e.g., ``province'', ``state'').

\begin{description}[itemsep=0.2em]
  \item[Q4] What type of area is Mexico City in Mexico?
  \item[Q6] New South Wales is an Australian what?
  \item[Q7] What type of subdivision is Alabama in the United States of America?
  \item[Q15] Which type of subdivision is Xinjiang in the People's Republic of China?
\end{description}

Graph-wise, these correspond to \emph{type edges} (e.g., \texttt{instanceOf}, \texttt{subclassOf}) plus optional conditioning on a containing unit. In abstract form, they recover a class $c$ such that $(v,\texttt{type},c)$, sometimes with an added constraint that $v$ is contained in a specified parent $p$ (e.g., $(v,\texttt{locatedIn},p)$). This family stresses correct disambiguation between \emph{entity} names and \emph{type} labels, and (when present) joining type reasoning with containment.

\subsubsection{\textbf{Selection Under Type and Location Constraints.}}

These queries invert the previous family: given a class (and sometimes a country), return an example entity.

\begin{description}[itemsep=0.2em]
  \item[Q5] Which one is a state in Australia?
  \item[Q9] Which federal district is located in the United States?
  \item[Q13] Which type of subdivision is found in England?
\end{description}

These are \emph{type-filtered selections} possibly combined with a containment predicate. For Q5/Q9, the canonical form is $\{v : (v,\texttt{type},\textsf{State}) \wedge (v,\texttt{locatedIn},\textsf{Australia})\}$ (similarly for \textsf{FederalDistrict} and \textsf{UnitedStates}). Q13 can be seen as a \emph{schema-level query} over the induced subgraph of England's administrative units: return the set (or a representative) of subdivision types $c$ such that $\exists v\, (v,\texttt{locatedIn},\textsf{England}) \wedge (v,\texttt{type},c)$. From the engine's perspective, this family stresses multi-predicate filtering (type $\wedge$ location), and for Q13, aggregation over types (distinct $c$).

\subsubsection{\textbf{Political Composition / Multi-Answer Part Lists.}}

These queries ask for a political unit and \emph{its other constituents}, yielding a small set of entities.

\begin{description}[itemsep=0.2em]
  \item[Q12] England is a constituent of which political unit, along with which other parts?
\end{description}

Graph-theoretically, Q12 composes (i) an upward containment step to find the parent political unit $u$ such that $(\textsf{England},\texttt{partOf},u)$, and (ii) a sibling expansion to enumerate other constituents $\{v : (v,\texttt{partOf},u)\}$, typically with a filter $v \neq \textsf{England}$. This is a minimal example of a \emph{two-stage join}: find a hub node $u$ and then fan out over its incident edges, stressing the system's ability to retain intermediate bindings (the found $u$) and reuse them across steps.

\section{USE CASES: DETAILED LISTINGS}
\label{sec:app:use-cases}

We provide detailed listings of use cases from Section~\ref{sec:usecases} (Listing~\ref{lst:e1}-\ref{lst:e5}).

\begin{figure*}[hbtp]
\begin{lstlisting}[label=lst:e1, caption={\textbf{Example use case of \schemenameS -- Aggregating early-stage drive assemblies per factory site.}}, basicstyle=\tt\small]
// User input question: 
|\rm\small\emph{How many base-tier drive assemblies do we produce at each factory site?}|

// The LLM Agent first constructs an appropriate tool call signature:
get_nodes(
  node_type = "DriveAssembly",
  filters = [{ "key": "assemblyTier", "operator": "=", "value": 0, "value_type": "number" }],
  group_by = "factoryCode",
  aggregations = [{ "grouping_type": "COUNT", "property": "*" }] )

// The Non-LLM Executor generates and executes a Cypher query:
MATCH (d:DriveAssembly) WHERE d.assemblyTier = 0 RETURN d.factoryCode AS factory, COUNT(*) AS assemblyCount

// The tool execution returns the following example data:
[{ "factory": "PLANT-EV-01", "assemblyCount": 829 },
{ "factory": "PLANT-EV-03", "assemblyCount": 3441 },
...]

// The LLM Agent replies to the user:
|\rm\small\emph{The number of early-tier drive assemblies (assembly tier 0) produced per factory site is:}|
|\rm\small\emph{- Factory PLANT-EV-01: 829 assemblies}|
|\rm\small\emph{- Factory PLANT-EV-03: 3441 assemblies}|
|\rm\small\emph{- ...}|
\end{lstlisting}
\end{figure*}

\begin{figure*}[hbtp]
\begin{lstlisting}[label=lst:e2, caption={\textbf{Example use case of \schemenameS -- Inspecting a manufacturing blueprint for a vehicle model.}}, basicstyle=\tt\small, escapechar=\#]
// User input request: 
#\rm\small\emph{Show me the unique manufacturing blueprint for vehicle model EV-X7.}#

// The LLM Agent first constructs an appropriate tool call signature:
get_unique_manufacturing_blueprints_for_model(model_name = "EV-X7", cost_type = "Plan")

// The Non-LLM Executor generates and executes a Cypher query:
MATCH (v:VehicleModel { name: "EV-X7" })
OPTIONAL MATCH (n1)-[rel1:INTEGRATED_IN|ASSEMBLED_AT|BUILT_AT]->(v)
OPTIONAL MATCH (n2)-[rel2:INTEGRATED_IN|ASSEMBLED_AT|BUILT_AT]->(n1)
RETURN v, n1, rel1, n2, rel2;
// Result of the query is then algorithmically processed to:
//  - get all manufacturing plans for VehicleModel
//  - identify unique blueprints
//  - build new graph aggregating properties of duplicate flows
// Tool returns built graph:
{
  "graph_handle": "graph_evx7",
  "nodes": [
    { "id": "EV-X7", "labels": ["VehicleModel"], "properties": { ... } },
    { "id": "BM-A1", "labels": ["BatteryModule"], "properties": { ... } },
    { "id": "DA-B7", "labels": ["DriveAssembly"], "properties": { ... } },
     ... ],
  "relationships": [
    { "source": "BM-A1", "target": "DA-B7", "type": "INTEGRATED_IN", "properties": { ... } },
    { "source": "DA-B7", "target": "EV-X7", "type": "INTEGRATED_IN", "properties": { ... } },
    ...]
}

// The UI also renders the blueprint as a graph visualization.
// The LLM Agent replies to the user:
#\rm\small\emph{The unique manufacturing blueprint for vehicle model EV-X7 includes:}#
#\rm\small\emph{* Main battery module: BM-A1 (primary energy source).}#
#\rm\small\emph{* A drive assembly: DA-B7, integrating BM-A1 and additional subsystems.}#
#\rm\small\emph{* One or more assembly lines and factory sites where the manufacturing stages occur.}#
#\rm\small\emph{The blueprint graph captures, for each node, the associated production window and cost under the 'Plan' profile.}#
\end{lstlisting}
\end{figure*}

\begin{figure*}[hbtp]
\begin{lstlisting}[label=lst:e3, caption={\textbf{Example use case of \schemenameS -- Multi-step reasoning and self-recovery from errors.}}, basicstyle=\tt\small]
// User input request:
|\rm\small\emph{At which factory site is the component with the highest market price manufactured?}|

// The LLM Agent first constructs an appropriate tool call signature:
get_nodes(
  node_type = "BatteryModule", aggregations = [{ "grouping_type": "MAX", "property": "marketPrice" }])

// The Non-LLM Executor generates and executes a Cypher query:
MATCH (b:BatteryModule) RETURN MAX(b.marketPrice) AS maxPrice

// The tool execution returns the following example data:
{ "maxPrice": 99.963895876119216 }

// The LLM Agent then realizes that the data does not identify the component itself. Thus, it generates another appropriate tool call signature:
get_nodes(node_type = "BatteryModule", order_by = "marketPrice", limit = 1, descending = true)

// The Non-LLM Executor generates and executes a Cypher query:
MATCH (b:BatteryModule) RETURN b ORDER BY b.marketPrice DESC LIMIT 1

// The tool execution returns the following example data:
{
  "moduleName": "BM-9003",
  "marketPrice": 99.963895876119216,
  "moduleKey": "BMK-9003",
  ...
}

// The LLM Agent then generates a final appropriate tool call signature:
get_factory_sites_for_module(module_name = "BM-9003")

// The Non-LLM Executor generates and executes a Cypher query:
MATCH (b:BatteryModule { name: "BM-9003" })-[:BUILT_AT]->(f:FactorySite)
RETURN DISTINCT f.siteName

// The tool execution returns the following example data:
{ "siteName": "Factory Alpha - Northern Division" }

// The LLM Agent replies to the user:
|\rm\small\emph{The battery module with the highest market price is manufactured at "Factory Alpha – Northern Division".}|
\end{lstlisting}
\end{figure*}

\begin{figure*}[hbtp]
\begin{lstlisting}[label=lst:e4, caption={\textbf{Example use case of \schemenameS -- Parallel tool calls and production plan comparison.}}, basicstyle=\tt\small]
// User input question: 
|\rm\small\emph{Check unique production plans for three vehicle models whose names start with 'B', and tell me which one uses the largest number of modules.}|

// The LLM Agent first constructs an appropriate tool call signature:
get_vehicle_models_by_prefix(name_prefix = "B", limit = 3)

// The Non-LLM Executor generates and executes a Cypher query:
MATCH (v:VehicleModel) WHERE v.name STARTS WITH "B" RETURN v.name LIMIT 3;

// The tool execution returns the following example data:
[
  { "modelName": "BX-985G9L", "internalName": "BX-985G9L ..." },
  { "modelName": "BE-2D2173", "internalName": "BE-2D2173 ..." },
  { "modelName": "BR-410PCE", "internalName": "BR-410PCE ..." }
]

// The LLM Agent then constructs a tool call signature for each returned name, e.g.,:
get_unique_production_plan_for_model(model_name = "BX-985G9L", cost_type = "Plan")

// The Non-LLM Executor generates and executes Cypher queries for each signature, e.g.:
MATCH (m:BatteryModule)-[:CONSUMED_IN*1..2]->(v:VehicleModel { name: "BX-985G9L" })
RETURN COUNT(DISTINCT m) AS moduleCount;

// The tool execution returns graph representations of the production plans for all three vehicle models

// The LLM Agent replies to the user:
|\rm\small\emph{Comparing the three production plans: BX-985G9L uses 14 distinct modules, BE-2D2173 uses 9 distinct modules, and BR-410PCE uses 11 distinct modules. Therefore, BX-985G9L is assembled from the largest number of modules.}|
\end{lstlisting}
\end{figure*}

\begin{figure*}[hbtp]
\begin{lstlisting}[label=lst:e5, caption={\textbf{Example use case of \schemenameS -- Self-correction and human-in-the-loop.}}, basicstyle=\tt\small]
// User input question: 
|\rm\small\emph{What is the highest module cost?}|

// The LLM Agent first constructs an appropriate tool call signature:
get_nodes(
  node_type = "BatteryModule", 
  aggregations = [{ "grouping_type": "MAX", "property": "unitCost" }], 
  limit = 1)

// The Non-LLM Executor generates and executes a graph query:
MATCH (m:BatteryModule) RETURN MAX(m.unitCost) AS maxCost

// The tool execution returns the following example data:
{ "maxCost": 99.419984557066762 }

// The LLM Agent replies to the user:
|\rm\small\emph{The highest battery module cost recorded in the database is approximately 99.42. If you would like, I can find the corresponding module for you.}|

// Follow-up user input:
|\rm\small\emph{Which module has this highest cost?}|

// The LLM Agent constructs an appropriate tool call signature:
get_nodes(
  node_type = "BatteryModule",
  filters = [{ "key": "unitCost", "operator": "=", "value": 99.42, "value_type": "number" }],
  limit = 1)

// The tool execution returns the following example data:
[] // No data found (float precision mismatch)

// The LLM Agent reacts appropriately and constructs another tool call signature:
get_nodes(node_type = "BatteryModule", limit = 1, order_by = "unitCost", descending = true)

// The Non-LLM Executor generates and executes a corresponding graph query:
MATCH (m:BatteryModule) RETURN m ORDER BY m.unitCost DESC LIMIT 1

// The tool execution returns the following example data:
{
  "moduleName": "MOD-X300R7",
  "unitCost": 99.419984557066762,
  "processTier": 1,
  "moduleKey": "MOD-7328272",
  ...
}

// The LLM Agent replies to the user:
|\rm\small\emph{The battery module with the highest cost ($\approx$ 99.42) is "MOD-X300R7". It belongs to process tier 1 and has the key MOD-7328272. It is used in downstream assembly lines and vehicle models.}|
\end{lstlisting}
\end{figure*}

\end{document}